\documentclass[aps,prb,showpacs,citeautoscript,superscriptaddress,twocolumn]{revtex4}

\usepackage{graphicx}
\usepackage{amsmath}
\usepackage{amssymb}
\usepackage{color}
\usepackage{subfigure}

\newcommand{\bea}{\begin{eqnarray*}}
\newcommand{\eea}{\end{eqnarray*}}
\newcommand{\bne}{\begin{equation*}}
\newcommand{\ede}{\end{equation*}}

\newcommand{\bnen}{\begin{equation}}
\newcommand{\eden}{\end{equation}}
\newcommand{\bean}{\begin{eqnarray}}
\newcommand{\eean}{\end{eqnarray}}
\newcommand{\bnsn}{\begin{subequations}}
\newcommand{\edsn}{\end{subequations}}

\newcommand{\bna}{\begin{array}}
\newcommand{\eda}{\end{array}}
\newcommand{\bnm}{\begin{enumerate}}
\newcommand{\edm}{\end{enumerate}}
\newcommand{\bni}{\begin{itemize}}
\newcommand{\edi}{\end{itemize}}

\renewcommand{\vec}[1]{\text{\boldmath{$ #1 $}}}

\newcommand{\beff}{\mathcal{B}}
\newcommand{\vbeff}{\vec{\mathcal{B}}}

\newcommand{\ket}[1]{| #1 \rangle}
\newcommand{\bra}[1]{\langle #1 |}

\begin{document}
\title{Shape-sensitive Pauli blockade in a bent
carbon nanotube}

\author{G\'abor Sz\'echenyi}
\affiliation{Institute of Physics, E\"otv\"os University, Budapest, Hungary}

\author{Andr\'as P\'alyi}
\affiliation{Institute of Physics, E\"otv\"os University, Budapest, Hungary}
\affiliation{MTA-BME Condensed Matter Research Group,
Budapest University of Technology and Economics, Budapest, Hungary}

\date{\today}

\begin{abstract}
Motivated by a recent experiment 
[F. Pei \emph{et al.}, Nat. Nanotech. {\bf 7}, 630 (2012)], we theoretically study
the Pauli blockade transport 
effect in a double quantum dot embedded in a bent carbon nanotube. 
We establish a model for Pauli blockade, taking into account
the strong g-factor anisotropy that is linked to the local orientation
of the nanotube axis in each quantum dot. 
We provide a set of conditions under which our model 
is approximately mapped to the spin-blockade model of Jouravlev and Nazarov
[O. N. Jouravlev and Y. V. Nazarov, Phys. Rev. Lett. {\bf 96}, 176804 (2006)].
The results we obtain for the magnetic anisotropy of the leakage current,
together with their qualitative geometrical explanation, 
provide a possible interpretation of previously unexplained
experimental results. 
Furthermore, we find that in a certain parameter range, 
the leakage current becomes highly sensitive to the shape of the tube,
and this sensitivity increases with increasing g-factor anisotropy.
This mutual dependence of the electron transport and the tube shape
allows for mechanical control of the leakage current,
and for characterization of the tube shape via measuring the leakage current. 
\end{abstract}


\maketitle

\section{Introduction}

\begin{figure}
\includegraphics[width=0.5\textwidth]{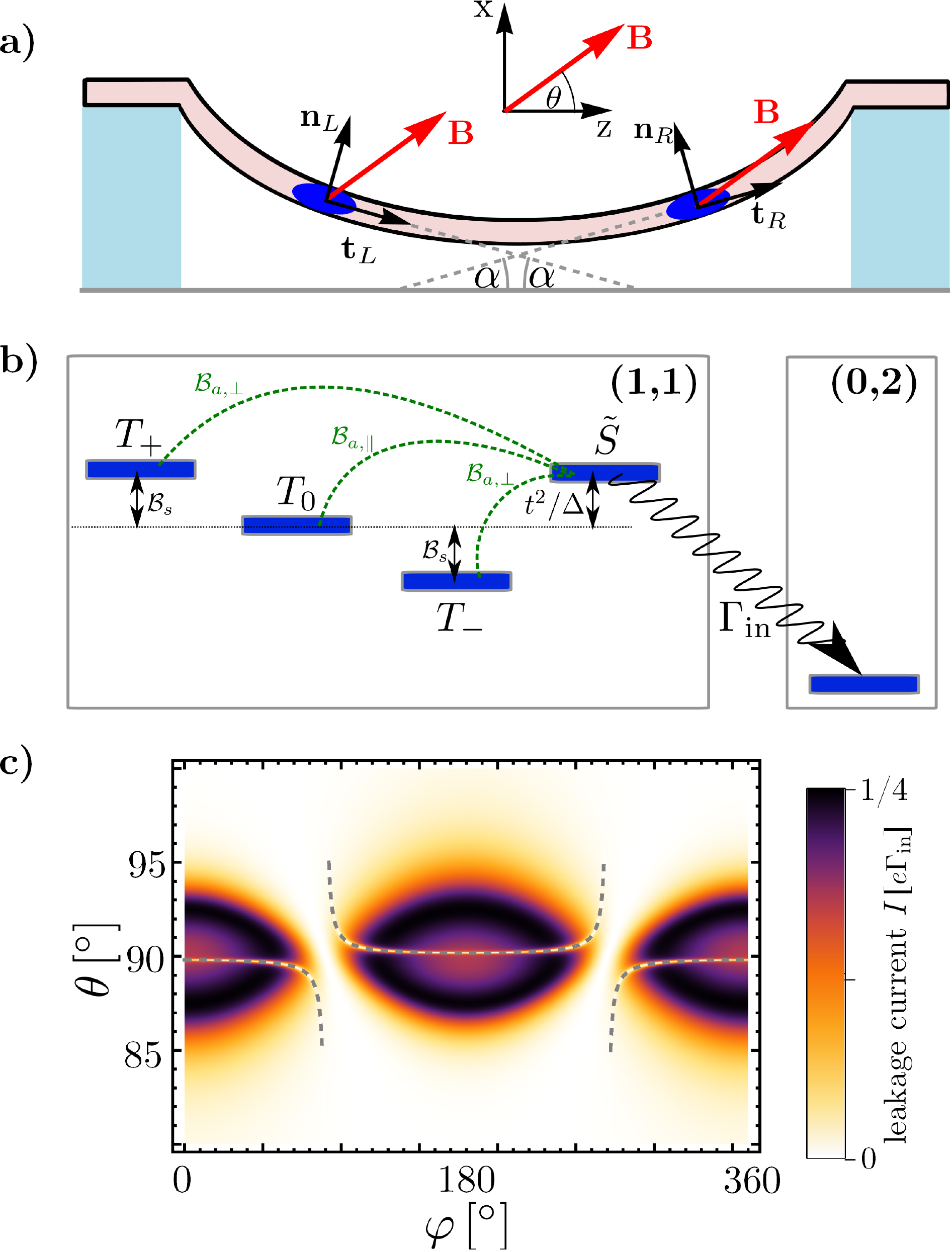}
\caption{\label{fig:setup} 
Pauli blockade in a bent carbon nanotube.
(a) Double quantum dot in a bent CNT in a homogeneous
magnetic field $\vec B$.
The direction of the magnetic field is characterized by 
the polar and azimuthal angles $\theta$ and $\varphi$. 
The geometry of the CNT is characterized by the
deflection angle $\alpha$. 
(b) Level diagram in the inelastic interdot 
tunneling regime [cf. Eq. \eqref{eq:hprime}].
Dashed lines represent matrix elements mixing the triplet
states with the singlet $\tilde S$. 
(c)
Theoretical result for the magnetic anisotropy of the leakage current
$I(\varphi,\theta)$ at $B=0.5$ T, showing
features similar to the measured data (Fig. S7 in the Supplementary Information of Ref. \onlinecite{FeiPei}).
Dashed grey line shows the position of the antiresonance
as described by the analytical formula Eq. \eqref{eq:antiresonance}.
Parameter values: $\alpha = 3^\circ$,
$g_\parallel = 32$, $E_{\tilde S} \equiv t^2/\Delta = 5\, \mu$eV,
$g_\perp^{(L)}=1.125$, $g_\perp^{(R)}=0.75$, $\gamma = 0$.
See Sec. \ref{sec:results} for more details.}
\end{figure}

Recent advances enable the fabrication of 
ultraclean individual carbon nanotubes (CNTs)
with exceptional electronic and mechanical quality\cite{Cao-cleancnt,Kuemmeth,Jespersen,Huttel-cntresonator,Steele-science,Steele-nn,Wu-stamping,FeiPei,Laird,Waissman,Benyamini}.
Transport experiments\cite{Laird-cm-review} in such devices
are aiming at, e.g., establishing strongly correlated electronic phases\cite{Deshpande,Pecker},
controlling the CNT's electronic and mechanical degrees of freedom
and their interactions\cite{Sazonova,Steele-science,Lassagne,Benyamini,Poot-review}, and
electron-spin-based quantum information processing\cite{Bulaev,ChurchillPRL,FlensbergMarcus,FeiPei,Laird}.

Characteristic of CNTs is the coexistence of mechanical flexibility
and strong spin-orbit interaction\cite{Ando,Kuemmeth,Jeong,Izumida}.
In combination with electrical confinement in CNT quantum dots (QDs), 
their interplay allows for strong spin-phonon coupling\cite{Bulaev,RudnerRashba,Palyi-spinphonon}, 
bend-induced and electrically controlled g-tensor modulation\cite{Lai},
electrically driven spin resonance\cite{FlensbergMarcus,Laird,Szechenyi-maximalrabi,Wang-control,Li-edsr,Osika-cnt}, 
spin-based motion sensing\cite{Ohm},
and mechanical readout of  spin-based quantum 
bits\cite{Struck-readout}.

In this work, we provide a theoretical description of a recently 
realized experimental setup\cite{FeiPei}, 
where the Pauli  blockade
transport effect was measured in a double QD (DQD) 
embedded in a bent CNT. 
A schematic of the setup is shown in Fig. \ref{fig:setup}a. 
In the Pauli blockade\cite{Ono-spinblockade,Jouravlev,Koppens-spinblockade,Buitelaar}, electronic transport through the serially coupled
DQD proceeds
via the (1,1) $\to$ (0,2) $\to$ (0,1) $\to$ (1,1) cycle of transitions,
where $(N_L,N_R)$ denotes the numbers of electrons in 
the neighboring QDs.

First, consider the case when the only internal degree of freedom of the
electrons is the spin, and hence the current flow is influenced by
the spin selection rules of the transitions of the transport cycle. 
This case is relevant for, e.g., III-V semiconductors\cite{Ono-spinblockade,Fransson,Jouravlev}, 
and CNTs with strong disorder\cite{Buitelaar,Chorley}. 
In
the absence of singlet-triplet mixing,
the (1,1) $\to$ (0,2) transition is forbidden for the triplet states
by Pauli's exclusion principle, hence the current is zero.
This blockade is lifted by spin perturbations causing singlet-triplet
mixing (e.g., spin-orbit interaction, hyperfine interaction, inhomogeneous 
magnetic field), inducing a nonzero leakage current.
In turn, measurement of the leakage current
can be used to characterize the spin Hamiltonian 
governing the current-carrying
electrons. 
The Pauli-blockade mechanism is also utilised for
qubit initialisation and readout in experiments\cite{Petta,Koppens-esr,Hanson-rmp}
demonstrating coherent control of few-electron quantum bits. 

In ultraclean CNT DQDs, the valley degree of freedom of the electrons 
and the large spin-orbit interaction play essential roles in 
Pauli blockade. 
In the  limit of vanishing valley mixing, and in the absence
of an external magnetic field, the ground-state doublet in
each QD is a Kramers pair, usually denoted by
$\ket{K\uparrow}$ and $\ket{K'\downarrow}$, 
with opposite spin orientation and different valley index.
The phenomenology of Pauli blockade, also called spin-valley blockade\cite{Palyi-hyperfine,Palyi-cnt-spinblockade}
or valley-spin blockade\cite{FeiPei} in this context, remains similar to the
spinful case: triplet-like two-electron states composed from 
$\ket{K\uparrow}$ and $\ket{K'\downarrow}$ block the current
in the absence of singlet-triplet mixing, and this blockade
can be lifted by spin- or valley perturbations acting differently in 
the two QDs.

Here, we present a model for the Pauli-blockade transport effect in a
DQD embedded in a bent CNT. 
We take into account
the strong g-factor anisotropy which is linked to the local orientation
of the nanotube axis in each QD (see Fig. \ref{fig:setup}a). 
We provide a set of conditions under which our model 
can be mapped to the spin-blockade model of Jouravlev and Nazarov\cite{Jouravlev}.
We calculate the dependence of the leakage current on the
orientation of the external magnetic field. 
The results we obtain, see e.g., Fig. \ref{fig:setup}c, 
provide a possible interpretation of previously unexplained
experimental results\cite{FeiPei}.
Furthermore, we find that in a certain parameter range, 
the leakage current becomes highly sensitive to the shape of the tube,
and this sensitivity increases with increasing g-factor anisotropy.
This mutual dependence of the electron transport and the tube shape
allows for mechanical control of the leakage current,
and for characterization of the tube shape via measuring the leakage current. 


\section{Magnetic anisotropy of the leakage current}
\label{sec:anisotropy}

Our aim in this work is to quantify the relation between the
Pauli-blockade 
leakage current and the system parameters, including the
shape of the CNT and the magnetic field vector.
A schematic of the  setup, along with the reference frame,
is shown in Fig. \ref{fig:setup}(a). 
The magnetic field vector is characterized by 
its magnitude $B$ and its usual polar $\theta$ and 
azimuthal $\varphi$ angles, $\vec B=B(\sin \theta \cos \varphi,\sin \theta \sin \varphi,\cos\theta)$.

For our analysis, 
experimental guidance is provided by the data of Ref. \onlinecite{FeiPei}.
There, an especially useful data set is presented in Fig. S7
of the Supplementary Information of Ref. \onlinecite{FeiPei} 
(to be referred to as S7 from now on).
In S7,  the
dependence of the leakage current on the magnetic field direction is 
plotted, 
in the case where
the (1,1)-(0,2) energy 
detuning (defined below) $\Delta\sim 1$ meV  and the magnetic field
strength $B = 0.5$ T are held fixed.
Importantly,
the detuning $\Delta$ was
chosen such that it exceeds the Zeeman splittings for any magnetic 
field direction in the angular range covered in S7. 
Furthermore, the current shown in S7 is presumably the result
of inelastic, e.g., phonon-emission-mediated, energetically downhill
(1,1) $\to$ (0,2) charge transitions (as depicted by the wavy 
arrow in Fig. \ref{fig:setup}b), 
as suggested by 
the detuning asymmetry of the current in the corresponding
data shown in Fig. 4a of Ref. \onlinecite{FeiPei}.

This measurement setting of S7 
simplifies the interpretation of the data: it suggests that
the rate $\Gamma_{\rm in}$
of the inelastic downhill (1,1) $\to$ (0,2)  tunneling process
is hardly sensitive to the magnetic field direction, hence
the observed field-direction dependence of 
 the leakage current is caused by the
 field-direction-induced variations of the four energy eigenstates of the (1,1)
charge configuration,
and not by variations of the tunnel rate $\Gamma_{\rm in}$.

The main features seen in S7 are as follows.
(E1) 
In a narrow range of $\theta \in [85^\circ,95^\circ]$, 
i.e., for magnetic field directions almost perpendicular to the CNT axis, 
the leakage current is much higher ($\sim 10$ pA)
than outside that range ($\lesssim 1$ pA). 
(E2) 
Apparently, the high-current region is defined by the
condition $| \theta - \pi/2| < A \cos \varphi$, where
$A \approx 5^\circ$.
(E3) 
The high-current regions are separated by low-current
gaps at $\varphi \approx 90^\circ$ and $\varphi \approx 270^\circ$.  
Two weaker features of S7:
(E4) 
There are two narrow lines of reduced current (antiresonances)
around $\theta \approx 90^\circ$,
approximately horizontal at $\varphi \approx 0$ ($\varphi \approx 180^\circ$)
but bending downward (upward) as $\varphi$ is moved away from $0$
($180^\circ$). 
(E5) 
There are two narrow lines of increased current (resonances),
at $(\varphi,\theta) \approx (0^\circ,98^\circ)$
[$(\varphi,\theta) \approx (180^\circ,82^\circ)$],
bending downward [upward] as $\varphi$ is moved away from $0$
[$180^\circ$].

Among the theoretical works
addressing few-electron physics in CNT DQDs\cite{Palyi-hyperfine,Weiss,Palyi-cnt-spinblockade,vonStecher,Reynoso1,Reynoso2,Kiss,Szechenyihotspot,Li-edsr}, 
Refs. \onlinecite{Palyi-cnt-spinblockade,Szechenyihotspot}
described the Pauli blockade transport effect
in the case of a 
\emph{straight} CNT.
In Ref. \onlinecite{Szechenyihotspot},
we found that the Pauli blockade can be 
lifted if the external magnetic field 
is perpendicular to the CNT axis.
That finding is in line with the experimental feature (E1)
seen in the bent CNT.
However, for the straight CNT,
the current is independent of the 
azimuth angle $\varphi$ of the field, because of the
 cylindrical symmetry of the straight geometry.
This is in contrast with the experimental features (E2)-(E5), 
which motivates the present study accounting for the bent shape 
of the CNT. 
The model we present will provide possible
explanations of the features (E1)-(E4) observed in S7,
as demonstrated by Fig. \ref{fig:setup}c.

\section{Model}

The setup, consisting of an electrostatically defined DQD in  
a bent CNT, is shown in Fig \ref{fig:setup}a.
In our model, the 
shape of the CNT is characterized by the unit vectors
$\vec t_{D}$ ($D =L,R$) along the local CNT axes in
the two QDs $L$ and $R$.
The reference frame (see Fig. \ref{fig:setup}{a}) is chosen such that these unit vectors
span the $x$-$z$ plane and are characterized by a single
angle parameter $\alpha$, which we refer to as the \emph{deflection angle}:
\begin{subequations}
\bean
\vec t_L &=& (-\sin \alpha, 0,\cos \alpha), \\
\vec t_R &=& (\sin \alpha, 0,\cos \alpha).
\eean
\end{subequations}
The deflection angle is assumed to be small\cite{Laird}, $\alpha \ll 1$.
We also introduce the unit vectors
$\vec n_L = (\cos \alpha, 0, \sin \alpha)$,
$\vec n_R = (\cos \alpha, 0, -\sin \alpha)$,
and $\vec n'_D = \vec t_D \times \vec n_D$,
see Fig. \ref{fig:setup}a.

\subsection{Single-electron Hamiltonian}

The $4\times 4$ Hamiltonian describing a single electron occupying the 
nominally fourfold (spin and valley) degenerate ground-state orbital 
of QD $D$
reads\cite{FlensbergMarcus,RudnerRashba}
\newcommand{\DD}{\Delta_{KK'}^{(D)}}
\newcommand{\gD}{\gamma^{(D)}}
\bean
\label{eq:HD}
H_D&=&
-\frac{\Delta_{\textrm{SO}}^{(D)}}{2}  \vec{ t}_D \cdot \vec{s}\, \tau_3 
+ \frac{\DD}{2} \left(
	\cos{\gD} \tau_1 + \sin{\gD} \tau_2
\right)
\nonumber
\\
&+& H_{B,D},\eean
where
\bean 
H_{B,D} &=&\frac{1}{2}g_s\mu_B \vec B \mbox{\boldmath{$s$}}
+\frac{1}{2}g_{v}^{(D)}\mu_B \vec t_D \cdot \vec B\, \tau_3.
\eean
Furthermore,
 $\vec s=(s_x,s_y,s_z)$ and $\boldsymbol\tau=(\tau_1,\tau_2,\tau_3)$ are the vectors of Pauli matrices in the spin and valley spaces, respectively, 
$\Delta_{\textrm{SO}}^{(D)}$ is the spin-orbit splitting in QD $D$, 
$\DD e^{i\gD}$ is the complex valley-mixing matrix element in QD $D$,
and the last two terms describe the Zeeman splittings, where $g_s$ ($g_v^{(D)}$)
  is the spin (orbital) g-factor (in QD $D$).  
We assume $\Delta_{\rm SO}^{(D)} >0$
and $g_v^{(D)} >0$.
In Eq. \eqref{eq:HD}, the unit matrices in spin ($s_0$) and valley ($\tau_0$)
space are suppressed.

The $8\times 8$ single-electron Hamiltonian of the DQD incorporates 
spin- and valley-conserving interdot tunnelling:
\bean
H_{\rm DQD} = H_L \eta_L + H_R \eta_R + H_{\rm tun},
\eean
where $\eta_{L/R} = (\eta_0\pm\eta_3)/2$,
\bean
H_{\rm tun} = \frac{t}{\sqrt 2} s_0 \tau_0 \eta_1,
\eean
and $\eta_{0,1,2,3}$ are the Pauli matrices
acting on the spatial degree of freedom $(L,R)$.
Furthermore, $t$ is real-valued.

At zero magnetic field $\vec B = 0$ and zero interdot 
tunnelling $t=0$, the energy eigenstates of $H_{\rm DQD}$ form
four Kramers doublets at energies $\pm \sqrt{[\Delta^{(D)}_{\rm SO}]^2+[\Delta^{(D)}_{KK'}]^2}$.
Here we focus on the low-energy doublets
in both dots, to be denoted by 
\begin{subequations}
\label{eq:kramerspair_old}
\bean
\ket{\tilde \Uparrow_D} &=& \cos{\frac{\chi^{(D)}}{2}}\ket{K \uparrow_D}-\sin{\frac{\chi^{(D)}}{2}}e^{i\gD}\ket{K' \uparrow_D},\\
 \ket{\tilde \Downarrow_D} &=&  \cos{\frac{\chi^{(D)}}{2}}\ket{K' \downarrow_D}-\sin{\frac{\chi^{(D)}}{2}}e^{-i\gD}\ket{K\downarrow_D},
\eean
\end{subequations}
where $K$ and $K'$ are the valley basis states,
$\uparrow_D$ ($\downarrow_D$) is
 the spin-up (spin-down) state  
 in QD $D$ with spin quantization axis $\vec{t}_D$,
and $\chi^{(D)} = \arctan\left(\DD/\Delta_{\rm SO}^{(D)}\right) \in [0,\pi/2[$.

In order to provide an approximate mapping
of our model of the CNT DQD to the model of Ref. \onlinecite{Jouravlev}
(see next subsection),
we introduce the following gauge-transformed 
states: 
\begin{subequations}
\label{eq:kramerspair}
\bean
\ket{\Uparrow_L} &=& e^{i\xi/2} \ket{\tilde \Uparrow_L},  \label{eq:ka}\\
\ket{\Downarrow_L} &=& e^{-i\xi/2} \ket{\tilde \Downarrow_L}, \label{eq:kb}\\
\ket{\Uparrow_R} &=&  e^{-i\xi/2} \ket{\tilde \Uparrow_R}, \label{eq:kc}\\
\ket{\Downarrow_R} &=& e^{i \xi/2} \ket{\tilde \Downarrow_R}, \label{eq:kd}
\eean
\end{subequations}
where
\bean
\label{eq:xi}
\xi = \arctan\left(
	\frac{\sin \gamma' \sin \frac{\chi^{(L)}}{2} \sin \frac{\chi^{(R)}}{2}}
	{\cos \frac{\chi^{(L)}}{2} \cos \frac{\chi^{(R)}}{2}
	+
	\cos \gamma' \sin \frac{ \chi^{(L)}}{2} \sin \frac{\chi^{(R)}}{2}}
\right),
\eean
and $\gamma' = \gamma^{(R)}-\gamma^{(L)}$.
The states defined in Eqs. \eqref{eq:ka} and \eqref{eq:kb}
[Eqs. \eqref{eq:kc} and \eqref{eq:kd}]
will be referred to as the 
 \emph{Kramers-qubit} basis states in QD $L$ [$R$].

\subsection{Low-energy single-electron Hamiltonian}

We assume conditions when only the four
lowest-energy single-particle energy levels  of
the DQD, i.e.,  
$\ket{\!\!\! \Uparrow_L}$,
$\ket{\!\!\! \Downarrow_L}$,
$\ket{\!\!\! \Uparrow_R}$,
 and
$\ket{\!\!\!\Downarrow_R}$,
participate in the (1,1)$\to$(0,2)$\to$(0,1)$\to$(1,1)
Pauli-blockade transport cycle.
The effects of interdot tunnelling and 
the external magnetic field are treated in 
first-order perturbation theory.
That is, we project the $8\times 8$ single-electron Hamiltonian
$H_{\rm DQD}$ 
to the
$4\times 4$ subspace spanned by
the four states above, i.e,
\bean
H'_{\rm DQD} \equiv P H_{\rm DQD} P,
\eean
where 
\bean
P = \sum_{D = L,R} \left(
	 \ket{\Uparrow_D} \bra{\Uparrow_D} + 
	 \ket{\Downarrow_D} \bra{\Downarrow_D}	 
	 \right).
\eean
The low-energy Hamiltonian $H'_{\rm DQD}$ 
 provides a good approximation for the dynamics 
as long as the spin and orbital
Zeeman splittings and the interdot tunnelling
$t$ are all much smaller than the energy splittings
$2 \sqrt{[\Delta^{(D)}_{\rm SO}]^2+[\Delta^{(D)}_{KK'}]^2}$
induced by spin-orbit interaction and valley mixing. 

Omitting a constant diagonal term in $H'_{\rm DQD}$,
it can be written as 
$H'_{\rm DQD} = H'_B + H'_{\rm tun}$.
As shown below, the homogeneous magnetic field is felt by
the Kramers-qubit in QD $D$ as a local effective magnetic field\cite{FlensbergMarcus,Szechenyihotspot} $\vbeff_D$.
This is made explicit by 
casting the low-energy magnetic Hamiltonian for the DQD in 
the following form:
\bean
\label{eq:hprimeb}
H'_{B} \equiv P H_B P =  \frac {1}{2} \left[
	\vec{\mathcal B}_L \vec \sigma_L 
	+ \vec{\mathcal B}_R \vec \sigma_R 
\right].
\eean
Here, $\vec{\sigma}_D$ is the vector of Pauli matrices
corresponding to the Kramers-qubit basis states in QD $D$, e.g., 
$\sigma_{L,3} = \ket{\Uparrow_L} \bra{\Uparrow_L} - \ket{\Downarrow_L} \bra{\Downarrow_L}$.
The  effective magnetic  field $\vec{\mathcal B}_D$ is related to the
external magnetic field $\vec B$ via
\begin{subequations}
\label{eq:beffcomponentsdqd}
\bean
\beff_{D1} &=&
-g_\perp^{(D)} \mu_B \textrm{Re}\left[(B_{Dn}+iB_{Dn'})e^{i(\gD+D \xi)}\right], \\
\beff_{D2} &=&
-g_\perp^{(D)} \mu_B \textrm{Im}\left[(B_{Dn}+iB_{Dn'})e^{i(\gD+D \xi)}\right], \\
\beff_{D3} &=& 
g_\parallel^{(D)} \mu_B B_{Dt}, 
\eean
\end{subequations}
where $D \in (L,R) \equiv (+1,-1)$, and
\begin{subequations}
\bean
g_\perp^{(D)}&=&g_s\sin{\chi^{(D)}},\\
g_\parallel^{(D)}&=&g_s+g_v^{(D)}\cos{\chi^{(D)}},
\eean
\end{subequations}
and we  introduced the projections of the
the external magnetic field
on the local coordinate axes via
\begin{subequations}
\bean
B_{Dt} &=& \vec t_D \cdot \vec B, \\
B_{Dn} &=& \vec n_D \cdot \vec B, \\
B_{Dn'} &=& \vec n'_D \cdot \vec B.
\eean
\end{subequations}
Henceforth, we will refer to 
$\beff_{D1}$ and $\beff_{D2}$ ($\beff_{D3}$)
as the transverse components (longitudinal component)
of the effective field,
and $g_\perp^{(D)}$ ($g_\parallel^{(D)}$) as the transverse
(longitudinal) g factor. 

We further define 
the symmetric $\vbeff_s = \frac 1 2 \left(	\vbeff_L	+ \vbeff_R
\right)$ and the antisymmetric $\vbeff_a = \frac 1 2 \left(	\vbeff_L - \vbeff_R\right)$ combinations  of the effective magnetic fields,
and the component $\vbeff_{a,\parallel}$ ($\vbeff_{a,\perp}$) 
of the antisymmetric combination that is parallel (perpendicular)
to $\vbeff_s$.

The single-electron tunnelling Hamiltonian in the low-energy subspace 
reads $H'_{\rm tun} \equiv P H_{\rm tun} P$.
We focus on cases where our model, 
at least approximately, can be mapped to 
that of Ref. \onlinecite{Jouravlev}.
To see when that can be done, let us recall a key feature of the
model of Ref. \onlinecite{Jouravlev}:
there is no tunneling between the (1,1) triplet states and
the (0,2) singlet state. This is ensured by the
fact that tunnelling is assumed to be spin-conserving,
i.e., in any spin basis, the single-electron 
tunnelling matrix elements are the same for the up-spin and down-spin
electrons.
In order to have the analogous feature, at least approximately, in our model, 
our tunneling Hamiltonian $H'_{\rm tun}$ 
should satisfy the following two conditions.
(i) Qubit-flip tunnelling should be much weaker
than qubit-conserving tunnelling,
i.e., 
$
|\bra{\Uparrow_L} H_{\rm tun} \ket{\Downarrow_R}|,
|\bra{\Uparrow_R} H_{\rm tun} \ket{\Downarrow_L}|
\ll
|\bra{\Uparrow_L} H_{\rm tun} \ket{\Uparrow_R}|,
|\bra{\Downarrow_L} H_{\rm tun} \ket{\Downarrow_R}|
$. 
(ii) 
The qubit-conserving tunnel amplitudes should be equal,
$
\bra{\Uparrow_L} H_{\rm tun} \ket{\Uparrow_R}
=
\bra{\Downarrow_L} H_{\rm tun} \ket{\Downarrow_R}
$.
Condition (i) is ensured if $\Delta^{(D)}_{\rm SO} \geq \Delta_{KK'}^{(D)}$ and
$\alpha \ll 1$, which we assume from now on. 
By explicit evaluation of the matrix elements of $H_{\rm tun}$, 
we find that the relation $\Delta^{(D)}_{\rm SO} \geq \Delta_{KK'}^{(D)}$
guarantees that the ratio of the qubit-flip and qubit-conserving
matrix elements fulfills
\bean
\frac{|\bra{\Uparrow_L} H_{\rm tun} \ket{\Downarrow_R}|}{|\bra{\Uparrow_L} H_{\rm tun} \ket{\Uparrow_R}|
 } \leq \sqrt{2} \tan \alpha  \ll 1,
\eean
hence the qubit-flip matrix elements can indeed be neglected to a good
approximation.
Condition (ii) is ensured by the gauge choice specified by Eqs. \eqref{eq:kramerspair} and \eqref{eq:xi}. (See also the discussion in Appendix C of Ref. \onlinecite{Li-edsr}.)
In fact, this gauge also guarantees that the qubit-conserving
tunnel amplitudes
$\bra{\Uparrow_L} H_{\rm tun} \ket{\Uparrow_R}$ and 
$\bra{\Downarrow_L} H_{\rm tun} \ket{\Downarrow_R}$
 are real-valued, but that is not essential.

%
%
%
%
%
%

\subsection{Two-electron Hamiltonian}

We consider the Pauli blockade occurring in the DQD in the
transport cycle $(0,1)\rightarrow(1,1)\rightarrow(0,2)\rightarrow(0,1)$.
We denote the energy detuning between the (1,1) and
(0,2) charge configurations by $\Delta$.
Together with the preceding assumptions, this provides the 
$5\times 5$
two-electron Hamiltonian 
\bean
H' = H'_B + t \left(
\ket{S}\bra{S_g} + \ket{S_g}\bra{S}
\right)
- \Delta \ket{S_g}\bra{S_g}
\eean 
Here $H'_B$ is the two-electron
generalisation of the single-electron
Hamiltonian in Eq. \eqref{eq:hprimeb}.
Furthermore, 
$\ket{S}$ [$\ket{S_g}$] denotes the singlet state
in the (1,1) [(0,2)] charge configuration,
formed from the local Kramers-qubit basis states. 
Note that through the preceding steps, 
we mapped the Pauli blockade problem in the bent 
CNT to the model of Jouravlev and Nazarov
\cite{Jouravlev}, originally developed to describe spin blockade
in GaAs in the presence of nuclear spins.

As discussed in Sec. \ref{sec:anisotropy}, 
the current shown in S7 is presumably the result of inelastic
(1,1) $\to$ (0,2) charge transitions.
Therefore we focus on the large-detuning 
case $\Delta \gg t$, and introduce the inelastic tunneling
rate $\Gamma_{\rm in}$ characterizing qubit-state-conserving
incoherent transitions
from the (1,1) to the (0,2) charge configuration. 
We eliminate the coherent tunnel
coupling $t$ from the Hamiltonian via perturbation theory,
resulting in `dressed' singlet states $\ket{\tilde S}$ and $\ket{\tilde S_g}$. 
The resulting $4\times 4$ Hamiltonian describing the (1,1) charge
configuration reads
\bean
\label{eq:hprime}
H'' = \left(\bna{cccc}
\beff_s & 0 & 0 & -\beff_{a,\perp}/\sqrt{2} \\
0 & 0 & 0 & \beff_{a,\parallel} \\
0 & 0 & -\beff_s & \beff_{a,\perp}/\sqrt{2} \\
-\beff_{a,\perp}/\sqrt{2} & \beff_{a,\parallel} & \beff_{a,\perp}/\sqrt{2} & E_{\tilde S} \\
\eda \right).
\eean
The basis  we use here is $T_+, T_0, T_-, \tilde S$,
where the triplet states are defined as usual,
but in a rotated qubit reference frame\cite{Jouravlev}
where the third axis is aligned with $\vbeff_s$ and
the first axis is aligned with $\vbeff_{a,\perp}$.
Furthermore, $E_{\tilde S} = t^2/\Delta$. 

The structure of the Hamiltonian $H''$ is visualised in Fig. \ref{fig:setup}b.
The symmetric combination of the effective magnetic fields
$\vbeff_s$ splits the three triplet states, whereas 
the antisymmetric combination $\vbeff_a$ is responsible for mixing
the triplet states with the dressed (1,1) singlet $\tilde S$. 
The Hamiltonian $H''$ allows us to identify special cases where
the current is zero\cite{Jouravlev,Danon-organic}.
If $\beff_{a,\perp} = 0$ ($\beff_{a,\parallel}=0$), then
$T_+$ and $T_-$  ($T_0$) decouple from $\tilde S$, 
and hence block the current. 
Another special case with zero current is $B_s = 0$: in this case,
a certain superposition of the three triplets forms a dark state
which is decoupled from $\tilde S$, and this dark state will block 
the current.

\subsection{Rate equation for the leakage current}

The leakage current is calculated as follows. 
First, we
diagonalize $H''$ to obtain its eigenstates 
$\ket{i}$ $(i=1,2,3,4)$.
Then, since qubit-flip tunnelling
is negligible, the (1,1) $\to$ (0,2) transition rate $\Gamma_i$ 
for each eigenstate $\ket{i}$
is assumed to be proportional to its $\ket{\tilde S}$ weight:
$\Gamma_i = \Gamma_{\rm in} \left| \langle \tilde S | i \rangle \right|^2$.
After reaching the (0,2) singlet state, 
one electron from QD $R$ exits to the drain, 
and one enters to QD $L$ from the source.
These steps are characterized by the filling rate $\Gamma_{\rm f}$ and
a corresponding probability $p_0$ of being either in the (0,2) or in the
(0,1) charge 
configuration. 
These considerations result in the following rate equations:
\begin{subequations}
\bean
\dot p_i &=& - \Gamma_i p_i + \frac 1 4 \Gamma_{\rm f} p_0 \ \ \ \ \ \ (i=1,2,3,4) \\
\dot p_0 &=& - \Gamma_{\rm f} p_0 + \sum_{i=1}^4 \Gamma_i p_i.
\eean
\end{subequations}
where $p_i$ ($i = 1,2,3,4$) is the occupation probability of the
(1,1) eigenstate $\ket{i}$.
Normalization condition $p_0 + \sum_{i=1}^4 p_i = 1$ also applies. 

We focus on the case when the bottleneck is the inelastic
interdot tunneling, i.e., $\Gamma_{\rm in} \ll \Gamma_{\rm f} $.
Then, 
the steady-state probabilities are $p_0 \approx 0$ and 
$p_i \approx \frac{1}{\Gamma_i \mathcal T }$ with 
$\mathcal T = \sum_{i=1}^4 \Gamma_i^{-1}$.
The steady-state leakage current is obtained
via $I= e \sum_{i=1}^4 p_i \Gamma_i = 4e/\mathcal{T}$.

\section{Results}
\label{sec:results}

In this section, we provide and discuss the results for the magnetic
anisotropy of the leakage current, and provide 
the corresponding geometrical interpretations based on the 
effective magnetic field vectors $\vbeff_{D}$. 

The effective magnetic fields $\vec{\mathcal B}_D$
given in Eq. \eqref{eq:beffcomponentsdqd} apparently depend explicitly on 
the complex phases $\gamma^{(D)}$ of the valley-mixing matrix elements
as well as the phase $\xi$ used for fixing the gauge.
However, it can be shown that the only combination of these parameters
that influences the current through the DQD is
$\gamma \equiv  \gamma^{(R)} - \gamma^{(L)} -  2\xi $.
Therefore, for our forthcoming results we specify the
value of the parameter $\gamma$.
Similarly, instead of specifying the values of 
the parameters $\Delta_{KK'}^{(D)}$, $\Delta_{\rm SO}^{(D)}$,
$g_v^{(D)}$ of our original model, 
we specify the  values of the derived parameters
$g_\perp^{(D)}$, $g_\parallel^{(D)}$, and $E_{\tilde S}$. 
For simplicity, we choose identical longitudinal g-factors in the 
two QDs, $g_\parallel\equiv g_\parallel^{(L)}=g_\parallel^{(R)}$. 

\subsection{Analytical results for the $E_{\tilde S}$ = 0 case}

In the special case $E_{\tilde{S}}= 0$, the following analytical 
result is obtained\cite{Jouravlev} for the leakage current:
\bean
\label{eq:jninelastic}
I = \frac 1 4 e \Gamma_{\rm in} (\vec {\hat {\mathcal B}}_L \times \vec{\hat{\mathcal B}}_R)^2,
\eean
where $\vec{\hat{\mathcal B}}_D = \vec{\mathcal B}_D / \mathcal{B}_D$.
This result implies that the maximal current
is $e\Gamma_{\rm in}/4$, and the current has this maximal value if
$\vec{\mathcal B}_L \perp  \vec{\mathcal B}_R$, i.e.,
$\vbeff_L \cdot \vbeff_R = 0$.

\begin{figure}
\includegraphics[width=0.5\textwidth]{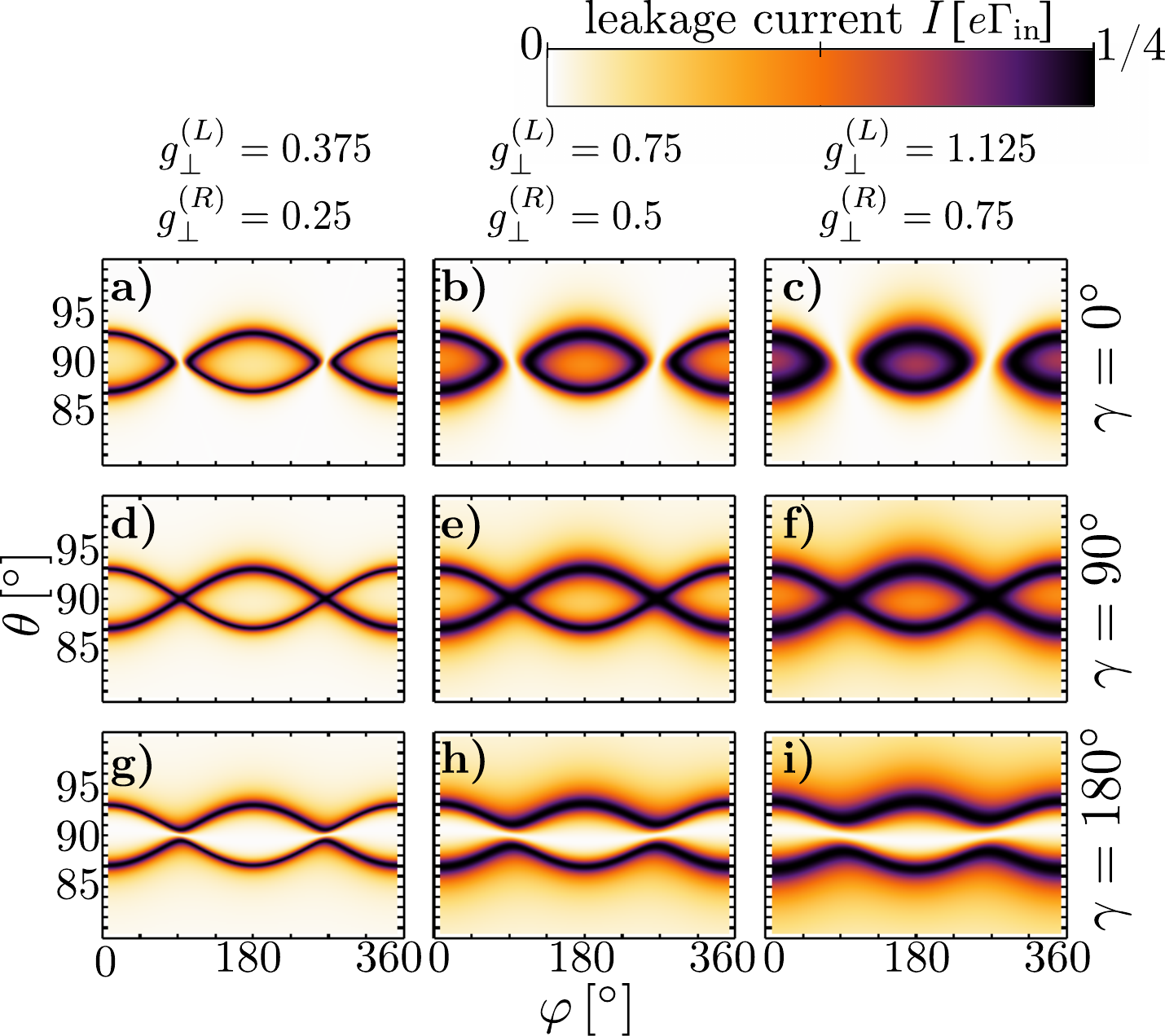}
\caption{\label{fig:valleymixing} 
Magnetic anisotropy of the leakage current: dependence on 
transverse g factors and the phase $\gamma$. 
The transverse g factors $g_\perp^{(L,R)}$ 
are shown at the top of each vertical block. 
The phase $\gamma$ 
is shown at the right end of each horizontal block.
Further parameters: $B=0.5$ T, $\alpha = 3^\circ$, $g_\parallel = 32$, $E_{\tilde S} = 0$.}
\end{figure}

The magnetic anisotropy of the leakage current \eqref{eq:jninelastic}
for a certain parameter set is shown in Fig. \ref{fig:valleymixing}a-f.
Focus on Figs. \ref{fig:valleymixing}a,d,g first, which corresponds to 
relatively small transverse g-factor values. 

To understand the results shown in Figs. \ref{fig:valleymixing}a,d,g 
it is instructive to consider the case of 
infinitesimally small transverse g factors 
(i.e., infinitesimally weak valley mixing).
In that limit, after expanding the maximal-current
condition $\vbeff_L \cdot \vbeff_R = 0$
up to second order in $\theta - \pi/2 \ll1$ and $\alpha \ll 1$, 
we obtain 
\bean
\label{eq:shapedependence}
\theta = \pi/2 \pm \alpha \cos \varphi,
\eean 
i.e., the maximal current $e \Gamma_{\rm in}/4$ is
flowing for magnetic field directions $(\varphi,\theta)$ 
fulfilling Eq. \eqref{eq:shapedependence}.
This makes sense: e.g., for $\varphi = 0$ and
$\theta = \pi/2 + \alpha$, the external magnetic field
is aligned with $\vec n_L$, hence the effective field $\vec{\mathcal B}_L$
is purely transversal, whereas $\vec{\mathcal B}_R$ is dominated
by its longitudinal component, i.e., these two vectors are
indeed perpendicular. 

Moreover, in the limit of infinitesimal transverse g factors,
current is finite only in the infinitesimal 
vicinity of the two maximal-current curves 
described by Eq. \eqref{eq:shapedependence}.
Otherwise the current is suppressed, for the following reason. 
If $\theta \neq \pi/2  \pm \alpha \cos \varphi$, then 
the longitudinal effective fields are finite in both dots and 
they dominate over the infinitesimally small
transverse effective fields. Then the effective field
vectors are almost parallel, hence, according
to Eq. \eqref{eq:jninelastic}, the current is almost zero.
This is exemplified in Figs. \ref{fig:valleymixing}a,d,g,
where the transverse g factors are set to
relatively small values, and 
hence the leakage current is significant only in 
the close vicinity of the lines given by Eq. \eqref{eq:shapedependence}. 

If the transverse g factors are
gradually increased, as shown in Figs \ref{fig:valleymixing}a,b,c,
then the narrow maximal-current lines become broader, 
eventually leading to an $I(\varphi,\theta)$
pattern that is very similar to feature (E2) of S7. 
[Note the remarkable similarity between Fig. \ref{fig:valleymixing}c 
and S7.]
Another way to phrase this is that in the $(\varphi,\theta)$
points in the vicinity of the lines $\theta = \pi/2 \pm \alpha \cos \varphi$, 
the leakage current increases with increasing transverse
g factors.
This effect is due to the fact that the increasing transverse
g factors increase the transverse components of the effective  fields
$\vbeff_D$
(cf. Eq. \eqref{eq:beffcomponentsdqd}), driving 
these fields away from their infinitesimal-transverse-g-factor limit where
 $\vbeff_L \parallel \vbeff_R$ and the current is zero.

Another feature seen in Figs. \ref{fig:valleymixing}a-c
is a low-current gap around $\varphi = 90^\circ$ and
$\varphi = 270^\circ$ between the
high-current regions.
This gap is getting larger as valley mixing is increased 
from Figs. \ref{fig:valleymixing}a to Figs. \ref{fig:valleymixing}c.
This gap is absent in Figs. \ref{fig:valleymixing}d-f,
where the phase  $\gamma$  is set to $\gamma = 90^\circ$. 
Also, the gap is not seen
in Figs. \ref{fig:valleymixing}g-i, 
where $\gamma = 180^\circ$. 
There, the 
$I(90^\circ,\theta)$ and $I(270^\circ,\theta)$ cuts
show high-current peaks for $\theta \approx 90^\circ \pm 1^\circ$.

These effects have straightforward geometrical interpretations based on
Eq. \eqref{eq:jninelastic}.
Furthermore, a quantitative description of these is
obtained if the maximal-current condition
$\vbeff_L \cdot \vbeff_R = 0$ is expanded up to second
order in $\theta - \pi/2 \ll1$, $\alpha \ll 1$ and 
$\frac{g_\perp^{(D)}}{g_\parallel } \ll 1$, yielding the maximal-current condition
\bean
\label{eq:maxcurrentcorrected}
\theta=\frac \pi 2 \pm\sqrt{(\alpha\cos{\varphi})^2-\frac{g_\perp^{(L)}g_\perp^{(R)}}{g_\parallel^2}\cos{\gamma}}.
\eean
Note that this refined version 
\eqref{eq:maxcurrentcorrected}
of Eq. \eqref{eq:shapedependence} 
depends on the 
phase $\gamma$.
The second term under the square root in Eq. \eqref{eq:maxcurrentcorrected},
proportional to $\cos \gamma$, 
accounts for the above described $\gamma$-dependent 
qualitative changes in Fig. \ref{fig:valleymixing}.



The above observations, together with the experimental data 
in S7, can be utilized to gain information on the 
 experimental setup of Ref. \onlinecite{FeiPei}.
In S7, the lines of maximal current are given approximately by
$\theta \approx 3^\circ \cos \varphi$, 
implying that the deflection angle is approximately $3^\circ$. 
Furthermore, feature (E3) implies that $0 \leq |\gamma| < \pi/2$;
 we use $\gamma = 0$ in the rest of this paper.
Finally, the fact that the leakage current at the center 
$(\varphi,\theta) = (0,\pi/2)$ of the
high-current region is almost as high as the maximal current
[at $(\varphi,\theta) \approx (0,3^\circ)$]
implies that the longitudinal and transverse effective field components
at $(\varphi,\theta) = (0,\pi/2)$ are similar in magnitude, i.e.
$g^{(D)}_\perp \sim g_\parallel \alpha$.

In conclusion, the analytical results for the $E_{\tilde S} = 0$ case
can describe the strong experimental features (E1), (E2) and (E3) 
if the model parameters are appropriately adjusted. 
In particular, the parameter set used in Fig. \ref{fig:valleymixing}c 
results in an $I(\varphi,\theta)$ pattern that is remarkably similar
to the experimental result S7. 
The weaker features (E4) and (E5) are not reproduced, 
motivating further study of the case $E_{\tilde S}>0$.

\subsection{Results for the $E_{\tilde S} > 0$ case}

\begin{figure}
\includegraphics[width=0.5\textwidth]{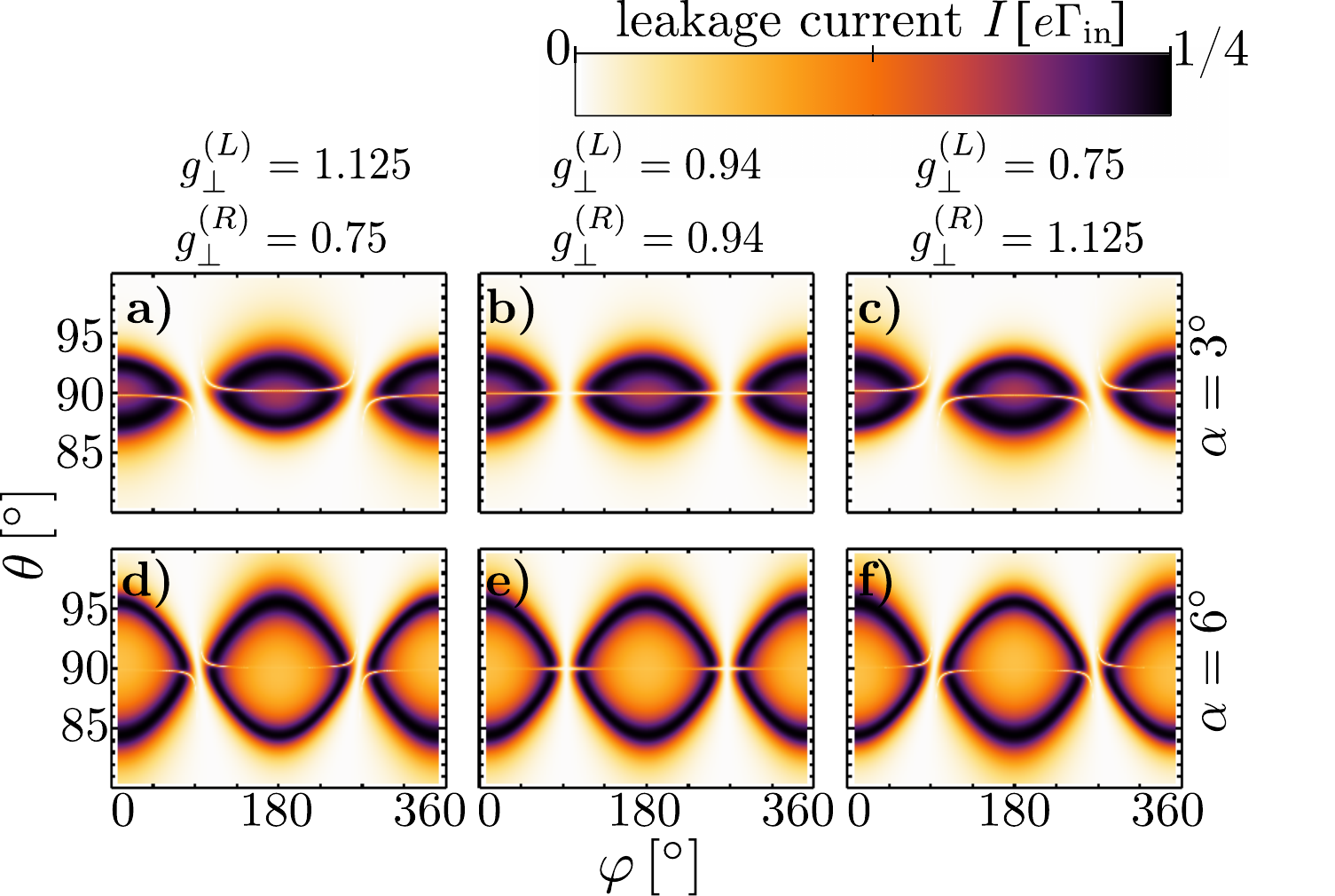}
\caption{\label{fig:angle} 
Magnetic anisotropy of the leakage current:
antiresonance and shape-sensitivity.
The transverse g factors $g_\perp^{(L,R)}$
are shown at the top of each vertical block. 
The deflection angle $\alpha$ characterizing the tube shape
is shown at the right end of each horizontal block.
Further parameters: $B=0.5$ T, $\gamma = 0$, 
$g_\parallel = 32$, $E_{\tilde S} = 5\  \mu$eV. }
\end{figure}

In this case, we diagonalize the Hamiltonian $H''$ numerically
to obtain the eigenstates $\ket{i}$. 
The magnetic anisotropy of the leakage current 
for specific choices of parameters close to 
the experimental values is shown in Fig. \ref{fig:angle}.
The narrow antiresonance (E4) appears on this plot, 
but the resonance (E5) does not. 
(Note that Figs. \ref{fig:setup}c and
\ref{fig:angle}a correspond to the same parameter set.)

The explanation of the antiresonance is as follows.
For certain magnetic field directions, the
effective fields in the two dots have the same magnitude
$\beff_L = \beff_R$. 
In that case, $\beff_{a,\parallel} = 0$, hence  the state $T_0$
decouples from the other three basis states in 
the $4\times 4$ Hamiltonian $H''$.
As $T_0$ is decoupled from $\tilde S$, it cannot decay to the (0,2)
singlet $\tilde S_g$, hence blocks the current flow and
therefore the leakage current vanishes in this case.

The shape of the antiresonance curve on the $\theta,\varphi$ plane
is described by the condition $\beff_L = \beff_R$, 
which, after linearizing in $\alpha \ll 1$ and $\theta -\pi/2 \ll 1$,
yields
\bnen
\label{eq:antiresonance}
\theta=\frac{\pi}{2}-\frac{\left(g_\perp^{(L)}\right)^2-\left(g_\perp^{(R)}\right)^2}{4g_\parallel^2}\frac{1}{\alpha\cos{\varphi}}.
\eden
Note that this result is consistent with the assumption $\theta - \pi/2 \ll 1$
only if the second term of the rhs of Eq. \eqref{eq:antiresonance}
is much smaller than 1. 
The analytical  result Eq. \eqref{eq:antiresonance}
is superimposed as a dashed line on Fig. 1c on the numerically 
obtained leakage-current density plot.
The analytical result follows closely the narrow low-current region
of the density plot. 
We remark that the function \eqref{eq:antiresonance} 
describing the antiresonance curve
is independent of the phase $\gamma$, which is a consequence of the
condition $\beff_L = \beff_R$ being insensitive to the directions
of the effective field vectors. 

Recall that the antiresonance appears in Fig. \eqref{fig:angle} 
because $E_{\tilde S}\equiv t^2/\Delta$ is set to a nonzero value. 
This implies that the visibility of the antiresonance
depends on $t$ and $\Delta$, and perhaps also
on further system parameters. 
To characterize this visibility, we analytically calculate
the leakage current in the vicinity of a 
given point $(\varphi_0,\theta_0)$ on the antiresonance curve.
We do this by taking into account the perturbative coupling
of the state $T_0$  to the state $\tilde S$ by
the small matrix element $\beff_{a,\perp}$,
see Eq. \eqref{eq:hprime}.
The leakage current is governed by the corresponding slow decay rate, 
and is evaluated using first-order perturbation theory in 
$\beff_{a,\perp}/E_{\tilde S}$.
This yields the following result:
\bean
\label{eq:visibility}
I(\varphi_0,\theta_0+\delta \theta)=
c \left(\frac{\Delta\mu_B
B}{t^2}\right)^2 \alpha^2\cos^2{\varphi_0} \, \delta \theta^2
e\Gamma_{\rm{in}},
\eean
where
\bean
c=16\frac{g_\parallel^4}{|g_\perp^{(L)}+g_\perp^{(R)}e^{i\gamma}|^2}.
\eean
Importantly, the prefactor of $\delta \theta^2$ in Eq. \eqref{eq:visibility}
decreases as $t^2/\Delta$ increases.
Therefore, increasing $t^2/\Delta$ 
increases the width of the antiresonance along the $\theta$ direction,
hence increases the visibility of the antiresonance. 
This is in line with our observations, i.e., with the appearance 
of the antiresonance upon increasing $t^2/\Delta$ from zero [Fig. \ref{fig:valleymixing}c] to a finite
value [Fig. \ref{fig:angle}a]. 

We note that an antiresonance effect similar to that in Fig. \ref{fig:angle}
has been discussed in Ref. \onlinecite{Jouravlev} for GaAs DQDs with
isotropic g-tensors and isotropic hyperfine interaction.
However, to our knowledge, such an antiresonance (called `stopping point'
in Ref. \onlinecite{Jouravlev}) has not been observed in GaAs DQDs.
The reason is probably that the external magnetic field vector
corresponding to a stopping point in GaAs depends on the nuclear 
spin configuration, and the latter typically changes significantly during 
a current measurement; hence the stopping points are averaged out
and the measured current appears to be a smooth function
of the magnetic field.  
Another type of stopping point, corresponding to the
condition $\vbeff_s = 0$, has been discussed in 
Ref. \onlinecite{Danon-organic}.

Figures \ref{fig:angle}a-f also demonstrate, in line with 
Eq. \eqref{eq:antiresonance}, that the $L$/$R$ asymmetry in the 
transverse g factors
is directly observable as the orientation of the 
antiresonance curve on the $I(\varphi,\theta)$ plot.
E.g., the antiresonance curve in the $90^\circ < \varphi < 270^\circ$
interval bends upwards (downwards) if
the transverse g factor is greater in QD $L$ ($R$),
and it is a flat line if the transverse g factors are 
equal.

\subsection{Shape sensitivity of the leakage current}

We use Fig. \ref{fig:angle} to demonstrate 
the dependence of the leakage current 
on the DQD's deflection angle $\alpha$.
The first line (Fig. \ref{fig:angle}a-c) shows  
the magnetic anisotropy of the leakage current
for $\alpha =3^\circ$, 
whereas the 
second line (Fig. \ref{fig:angle}d-f) shows that
for $\alpha = 6^\circ$. 
As predicted by Eq. \eqref{eq:shapedependence},
the region of maximal current is focused around
the lines $\theta = \pi/2 \pm  \alpha \cos \varphi$.
It is clear from Eq. \eqref{eq:antiresonance}, 
although less obvious from Fig. \ref{fig:angle},
that the antiresonance line moves as
the angle $\alpha$ is changed. 

\begin{figure}
\includegraphics[width=0.5\textwidth]{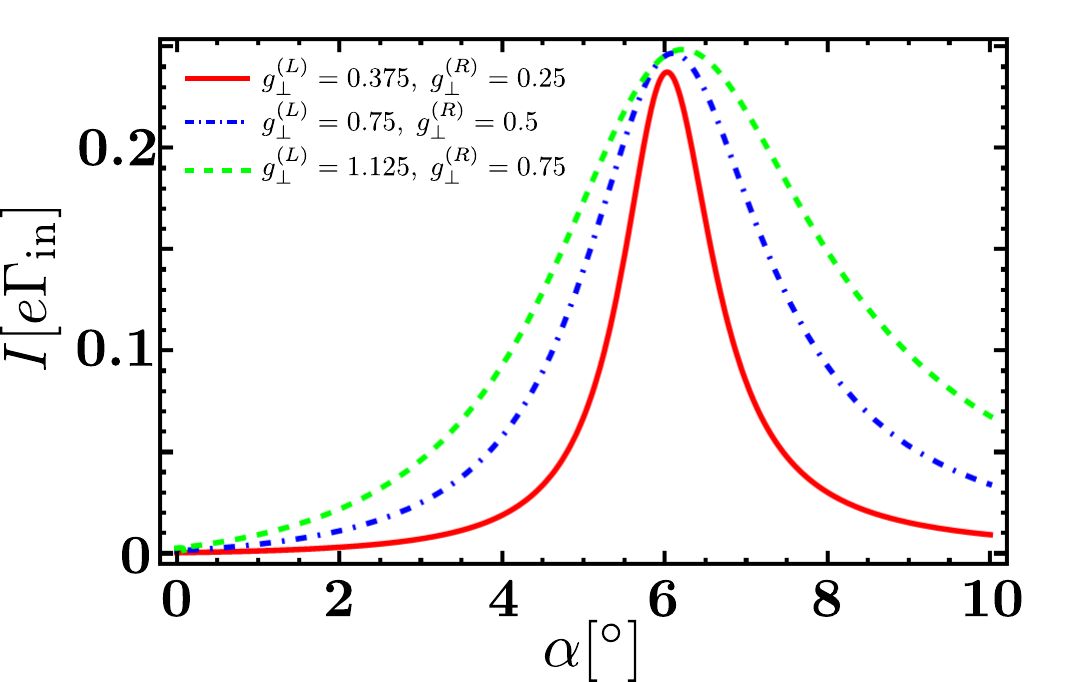}
\caption{\label{fig:shape} 
Shape-dependent leakage current in a bent carbon nanotube
double quantum dot. 
The three curves correspond to 
three different settings of the transverse g factors. 
By decreasing the transverse g factors, the peak width decreases, i.e., 
the current becomes more sensitive to the nanotube shape.
Parameter values: $B=0.5$ T, $\gamma = 0$, 
$g_\parallel = 32$, $E_{\tilde S} = 5\  \mu$eV. }
\end{figure}

In Fig. \ref{fig:shape}, we 
show how the leakage current depends on the
deflection angle $\alpha$ characterizing the shape of the CNT for
a fixed magnetic field.
To this end, we pick the  points $(\varphi,\theta) = (180^\circ,96^\circ)$
in Figs. \ref{fig:angle}a d,
and plot the leakage current for this magnetic field orientation
 as the deflection angle is varied continuously.
This is shown as the green dashed line in Fig. \ref{fig:shape},
which displays a broad peak around $\alpha = 6^\circ$.
The other two lines correspond to smaller transverse g factors
(i.e., smaller valley-mixing matrix elements or larger spin-orbit
splittings), 
resulting in narrowed current peaks.

The results shown in Fig. \ref{fig:shape} suggest that the \emph{in situ}
changes in the shape of the CNT could in principle be monitored by measuring
the current flowing through the embedded DQD. 
Static variations in $\alpha$ with respect to a reference 
value (`working point') $\alpha_0$ could be effectively detected if the 
slope of the $I(\alpha)$ curve is large at the reference value $\alpha_0$.
For example, if the system is described by the solid red curve of
Fig. \ref{fig:shape}, then, e.g., $\alpha_0 \approx 5.5^\circ$
is a good working point.
Dynamical variation of $\alpha$, due to, 
e.g., external driving of a
flexural phonon mode of the CNT\cite{Sazonova,Steele-science,Lassagne}, 
could also be detected as long as
its frequency is well below the tunnel rate $\Gamma_{\rm in}$. 
In that case, the working point $\alpha_0$ should be chosen such that
the second derivative of $I(\alpha)$ is large at $\alpha_0$.
Using the example of the solid red curve in Fig. \ref{fig:shape}, 
$\alpha_0 \approx 6^\circ$ is a good operating point. 
The large second derivative ensures that the time-averaged current
will be highly sensitive to the time-dependent variation of $\alpha$:
if $\alpha(t) = \alpha_0 + \delta \alpha \sin\omega t$,
then the time-averaged current is
$I_{\rm avg} \approx \frac{1}{T}\int_0^Tdt  I(\alpha(t))
\approx I(\alpha_0) + \frac{1}{4} (\delta \alpha)^2\left. \frac{d^2I}{d\alpha^2} \right|_{\alpha = \alpha_0} $.
These considerations together with Fig. \ref{fig:shape} imply that
the efficiency of the measurement of the static deflection angle and
its time-dependent variation improves if the transverse g factors are
decreased or the longitudinal g factors are 
increased.

The above-discussed
principle of detecting dynamical variations of the CNT deflection
 is similar to the one used in the experiments of Refs. \onlinecite{Huttel-cntresonator,Steele-science}. 
 There, the detection scheme was based on Coulomb-blockade peaks, and the deflection-current relation was induced
 by a capacitive mechanism. 
 In contrast, here the peak in $I(\alpha)$ arises because
 Pauli blockade is lifted, and the deflection-current relation 
 is due to the deflection-induced changes in the spin Hamiltonian.

\section{Conclusion}

We studied the dependence of the 
Pauli-blockade leakage current on 
the magnetic field direction in a DQD
embedded in a bent CNT,
and compared our results
to a recent experiment. 
The model we use reproduces a number of 
previously unexplained experimental features 
[see (E1)-(E4) of Sec. \ref{sec:anisotropy}].
We demonstrate that the leakage current is sensitive
to the shape of the CNT, and this sensitivity increases if
the ratio of the longitudinal and transverse g factors 
increases.
In principle, 
this sensitivity allows for mechanical control of the leakage current,
and a characterization of the tube shape via measuring the leakage current. 
For a recent experiment, we use our model to deduce a deflection angle 
of 3$^\circ$ from the measured magnetic anisotropy of the 
leakage current. 

Our model does not provide explanation for the
weak resonances (E5) seen in the experiment. 
There are a number of potential future extensions of
the present theory: 
accounting for (i) different longitudinal g-factors in the two dots, 
(ii) qubit-flip interdot tunneling, 
(iii) the n-p character of the double dot\cite{FeiPei,Li-edsr}, 
(iv) valley-mixing character of the electron-electron interaction\cite{Secchi-intervalley}, 
(v) Wigner-molecule physics\cite{Deshpande,Wunsch,SecchiRontani,vonStecher,Pecker,SecchiRontani2,CavaliereWigner}, etc.
We believe that incorporating these mechanisms 
would render the model more accurate quantitatively, 
and might also allow for an explanation of the observed resonance.

\begin{acknowledgments}
We acknowledge funding from 
the EU Marie Curie Career Integration Grant CIG-293834 (CarbonQubits),
the OTKA Grant PD 100373, 
the OTKA Grant 108676, and the EU ERC Starting Grant 
CooPairEnt 258789.
A.~ P.~ is supported by the 
J\'anos Bolyai Scholarship of the Hungarian Academy of Sciences.
\end{acknowledgments}

\bibliography{bend}

\begin{thebibliography}{55}
\expandafter\ifx\csname natexlab\endcsname\relax\def\natexlab#1{#1}\fi
\expandafter\ifx\csname bibnamefont\endcsname\relax
  \def\bibnamefont#1{#1}\fi
\expandafter\ifx\csname bibfnamefont\endcsname\relax
  \def\bibfnamefont#1{#1}\fi
\expandafter\ifx\csname citenamefont\endcsname\relax
  \def\citenamefont#1{#1}\fi
\expandafter\ifx\csname url\endcsname\relax
  \def\url#1{\texttt{#1}}\fi
\expandafter\ifx\csname urlprefix\endcsname\relax\def\urlprefix{URL }\fi
\providecommand{\bibinfo}[2]{#2}
\providecommand{\eprint}[2][]{\url{#2}}

\bibitem[{\citenamefont{Pei et~al.}(2012)\citenamefont{Pei, Laird, Steele, and
  Kouwenhoven}}]{FeiPei}
\bibinfo{author}{\bibfnamefont{F.}~\bibnamefont{Pei}},
  \bibinfo{author}{\bibfnamefont{E.~A.} \bibnamefont{Laird}},
  \bibinfo{author}{\bibfnamefont{G.~A.} \bibnamefont{Steele}},
  \bibnamefont{and} \bibinfo{author}{\bibfnamefont{L.~P.}
  \bibnamefont{Kouwenhoven}}, \bibinfo{journal}{Nat. Nanotech.}
  \textbf{\bibinfo{volume}{7}}, \bibinfo{pages}{630 } (\bibinfo{year}{2012}).

\bibitem[{\citenamefont{Cao et~al.}(2005)\citenamefont{Cao, Wang, and
  Dai}}]{Cao-cleancnt}
\bibinfo{author}{\bibfnamefont{J.}~\bibnamefont{Cao}},
  \bibinfo{author}{\bibfnamefont{Q.}~\bibnamefont{Wang}}, \bibnamefont{and}
  \bibinfo{author}{\bibfnamefont{H.~J.} \bibnamefont{Dai}},
  \bibinfo{journal}{Nat. Mater.} \textbf{\bibinfo{volume}{4}},
  \bibinfo{pages}{745} (\bibinfo{year}{2005}).

\bibitem[{\citenamefont{Kuemmeth et~al.}(2008)\citenamefont{Kuemmeth, Ilani,
  Ralph, and McEuen}}]{Kuemmeth}
\bibinfo{author}{\bibfnamefont{F.}~\bibnamefont{Kuemmeth}},
  \bibinfo{author}{\bibfnamefont{S.}~\bibnamefont{Ilani}},
  \bibinfo{author}{\bibfnamefont{D.~C.} \bibnamefont{Ralph}}, \bibnamefont{and}
  \bibinfo{author}{\bibfnamefont{P.~L.} \bibnamefont{McEuen}},
  \bibinfo{journal}{Nature} \textbf{\bibinfo{volume}{452}},
  \bibinfo{pages}{448} (\bibinfo{year}{2008}).

\bibitem[{\citenamefont{Jespersen et~al.}(2011)\citenamefont{Jespersen,
  Grove-Rasmussen, Paaske, Muraki, Fujisawa, Nyg�rd, and
  Flensberg}}]{Jespersen}
\bibinfo{author}{\bibfnamefont{T.~S.} \bibnamefont{Jespersen}},
  \bibinfo{author}{\bibfnamefont{K.}~\bibnamefont{Grove-Rasmussen}},
  \bibinfo{author}{\bibfnamefont{J.}~\bibnamefont{Paaske}},
  \bibinfo{author}{\bibfnamefont{K.}~\bibnamefont{Muraki}},
  \bibinfo{author}{\bibfnamefont{T.}~\bibnamefont{Fujisawa}},
  \bibinfo{author}{\bibfnamefont{J.}~\bibnamefont{Nyg�rd}}, \bibnamefont{and}
  \bibinfo{author}{\bibfnamefont{K.}~\bibnamefont{Flensberg}},
  \bibinfo{journal}{Nat. Phys} \textbf{\bibinfo{volume}{7}},
  \bibinfo{pages}{348} (\bibinfo{year}{2011}).

\bibitem[{\citenamefont{H\"uttel et~al.}(2009)\citenamefont{H\"uttel, Steele,
  Witkamp, Poot, Kouwenhoven, and van~der Zant}}]{Huttel-cntresonator}
\bibinfo{author}{\bibfnamefont{A.~K.} \bibnamefont{H\"uttel}},
  \bibinfo{author}{\bibfnamefont{G.~A.} \bibnamefont{Steele}},
  \bibinfo{author}{\bibfnamefont{B.}~\bibnamefont{Witkamp}},
  \bibinfo{author}{\bibfnamefont{M.}~\bibnamefont{Poot}},
  \bibinfo{author}{\bibfnamefont{L.~P.} \bibnamefont{Kouwenhoven}},
  \bibnamefont{and} \bibinfo{author}{\bibfnamefont{H.~S.~J.}
  \bibnamefont{van~der Zant}}, \bibinfo{journal}{Nano Lett.}
  \textbf{\bibinfo{volume}{9}}, \bibinfo{pages}{2547} (\bibinfo{year}{2009}).

\bibitem[{\citenamefont{Steele et~al.}(2009{\natexlab{a}})\citenamefont{Steele,
  H\"uttel, Witkamp, Poot, Meerwaldt, Kouwenhoven, and van~der
  Zant}}]{Steele-science}
\bibinfo{author}{\bibfnamefont{G.~A.} \bibnamefont{Steele}},
  \bibinfo{author}{\bibfnamefont{A.~K.} \bibnamefont{H\"uttel}},
  \bibinfo{author}{\bibfnamefont{B.}~\bibnamefont{Witkamp}},
  \bibinfo{author}{\bibfnamefont{M.}~\bibnamefont{Poot}},
  \bibinfo{author}{\bibfnamefont{H.~B.} \bibnamefont{Meerwaldt}},
  \bibinfo{author}{\bibfnamefont{L.~P.} \bibnamefont{Kouwenhoven}},
  \bibnamefont{and} \bibinfo{author}{\bibfnamefont{H.~S.~J.}
  \bibnamefont{van~der Zant}}, \bibinfo{journal}{Science}
  \textbf{\bibinfo{volume}{325}}, \bibinfo{pages}{1103}
  (\bibinfo{year}{2009}{\natexlab{a}}).

\bibitem[{\citenamefont{Steele et~al.}(2009{\natexlab{b}})\citenamefont{Steele,
  Gotz, and Kouwenhoven}}]{Steele-nn}
\bibinfo{author}{\bibfnamefont{G.~A.} \bibnamefont{Steele}},
  \bibinfo{author}{\bibfnamefont{G.}~\bibnamefont{Gotz}}, \bibnamefont{and}
  \bibinfo{author}{\bibfnamefont{L.~P.} \bibnamefont{Kouwenhoven}},
  \bibinfo{journal}{Nature Nanotech.} \textbf{\bibinfo{volume}{4}},
  \bibinfo{pages}{363} (\bibinfo{year}{2009}{\natexlab{b}}).

\bibitem[{\citenamefont{Wu et~al.}(2010)\citenamefont{Wu, Liu, and
  Zhong}}]{Wu-stamping}
\bibinfo{author}{\bibfnamefont{C.~C.} \bibnamefont{Wu}},
  \bibinfo{author}{\bibfnamefont{C.~H.} \bibnamefont{Liu}}, \bibnamefont{and}
  \bibinfo{author}{\bibfnamefont{Z.}~\bibnamefont{Zhong}},
  \bibinfo{journal}{Nano Lett.} \textbf{\bibinfo{volume}{10}},
  \bibinfo{pages}{1032} (\bibinfo{year}{2010}).

\bibitem[{\citenamefont{Laird et~al.}(2013)\citenamefont{Laird, Pei, and
  Kouwenhoven}}]{Laird}
\bibinfo{author}{\bibfnamefont{E.~A.} \bibnamefont{Laird}},
  \bibinfo{author}{\bibfnamefont{F.}~\bibnamefont{Pei}}, \bibnamefont{and}
  \bibinfo{author}{\bibfnamefont{L.~P.} \bibnamefont{Kouwenhoven}},
  \bibinfo{journal}{Nat. Nanotech.} \textbf{\bibinfo{volume}{8}},
  \bibinfo{pages}{565} (\bibinfo{year}{2013}).

\bibitem[{\citenamefont{Waissman et~al.}(2013)\citenamefont{Waissman, Honig,
  Pecker, Benyamini, Hamo, and Ilani}}]{Waissman}
\bibinfo{author}{\bibfnamefont{J.}~\bibnamefont{Waissman}},
  \bibinfo{author}{\bibfnamefont{M.}~\bibnamefont{Honig}},
  \bibinfo{author}{\bibfnamefont{S.}~\bibnamefont{Pecker}},
  \bibinfo{author}{\bibfnamefont{A.}~\bibnamefont{Benyamini}},
  \bibinfo{author}{\bibfnamefont{A.}~\bibnamefont{Hamo}}, \bibnamefont{and}
  \bibinfo{author}{\bibfnamefont{S.}~\bibnamefont{Ilani}},
  \bibinfo{journal}{Nature Nanotech.} \textbf{\bibinfo{volume}{8}},
  \bibinfo{pages}{569} (\bibinfo{year}{2013}).

\bibitem[{\citenamefont{Benyamini et~al.}(2014)\citenamefont{Benyamini, Hamo,
  Kusminskiy, von Oppen, and Ilani}}]{Benyamini}
\bibinfo{author}{\bibfnamefont{A.}~\bibnamefont{Benyamini}},
  \bibinfo{author}{\bibfnamefont{A.}~\bibnamefont{Hamo}},
  \bibinfo{author}{\bibfnamefont{S.~V.} \bibnamefont{Kusminskiy}},
  \bibinfo{author}{\bibfnamefont{F.}~\bibnamefont{von Oppen}},
  \bibnamefont{and} \bibinfo{author}{\bibfnamefont{S.}~\bibnamefont{Ilani}},
  \bibinfo{journal}{Nature Physics} \textbf{\bibinfo{volume}{10}},
  \bibinfo{pages}{151} (\bibinfo{year}{2014}).

\bibitem[{\citenamefont{Laird et~al.}()\citenamefont{Laird, Kuemmeth, Steele,
  Grove-Rasmussen, Nygard, Flensberg, and Kouwenhoven}}]{Laird-cm-review}
\bibinfo{author}{\bibfnamefont{E.}~\bibnamefont{Laird}},
  \bibinfo{author}{\bibfnamefont{F.}~\bibnamefont{Kuemmeth}},
  \bibinfo{author}{\bibfnamefont{G.}~\bibnamefont{Steele}},
  \bibinfo{author}{\bibfnamefont{K.}~\bibnamefont{Grove-Rasmussen}},
  \bibinfo{author}{\bibfnamefont{J.}~\bibnamefont{Nygard}},
  \bibinfo{author}{\bibfnamefont{K.}~\bibnamefont{Flensberg}},
  \bibnamefont{and} \bibinfo{author}{\bibfnamefont{L.~P.}
  \bibnamefont{Kouwenhoven}}, \bibinfo{note}{arXiv:1403:6113 (unpublished)}.

\bibitem[{\citenamefont{Deshpande and Bockrath}(2008)}]{Deshpande}
\bibinfo{author}{\bibfnamefont{V.~V.} \bibnamefont{Deshpande}}
  \bibnamefont{and} \bibinfo{author}{\bibfnamefont{M.}~\bibnamefont{Bockrath}},
  \bibinfo{journal}{Nature Physics} \textbf{\bibinfo{volume}{4}},
  \bibinfo{pages}{314} (\bibinfo{year}{2008}).

\bibitem[{\citenamefont{Pecker et~al.}(2013)\citenamefont{Pecker, Kuemmeth,
  Secchi, Rontani, Ralph, McEuen, and Ilani}}]{Pecker}
\bibinfo{author}{\bibfnamefont{S.}~\bibnamefont{Pecker}},
  \bibinfo{author}{\bibfnamefont{F.}~\bibnamefont{Kuemmeth}},
  \bibinfo{author}{\bibfnamefont{A.}~\bibnamefont{Secchi}},
  \bibinfo{author}{\bibfnamefont{M.}~\bibnamefont{Rontani}},
  \bibinfo{author}{\bibfnamefont{D.~C.} \bibnamefont{Ralph}},
  \bibinfo{author}{\bibfnamefont{P.~L.} \bibnamefont{McEuen}},
  \bibnamefont{and} \bibinfo{author}{\bibfnamefont{S.}~\bibnamefont{Ilani}},
  \bibinfo{journal}{Nature Physics} \textbf{\bibinfo{volume}{9}},
  \bibinfo{pages}{576} (\bibinfo{year}{2013}).

\bibitem[{\citenamefont{Sazonova et~al.}(2004)\citenamefont{Sazonova, Yaish,
  \"Ust\"unel, Roundy, Arias, and McEuen}}]{Sazonova}
\bibinfo{author}{\bibfnamefont{V.}~\bibnamefont{Sazonova}},
  \bibinfo{author}{\bibfnamefont{Y.}~\bibnamefont{Yaish}},
  \bibinfo{author}{\bibfnamefont{H.}~\bibnamefont{\"Ust\"unel}},
  \bibinfo{author}{\bibfnamefont{D.}~\bibnamefont{Roundy}},
  \bibinfo{author}{\bibfnamefont{T.~A.} \bibnamefont{Arias}}, \bibnamefont{and}
  \bibinfo{author}{\bibfnamefont{P.~L.} \bibnamefont{McEuen}},
  \bibinfo{journal}{Nature} \textbf{\bibinfo{volume}{431}},
  \bibinfo{pages}{284} (\bibinfo{year}{2004}).

\bibitem[{\citenamefont{Lassagne et~al.}(2009)\citenamefont{Lassagne,
  Tarakanov, Kinaret, Garcia-Sanchez, and Bachtold}}]{Lassagne}
\bibinfo{author}{\bibfnamefont{B.}~\bibnamefont{Lassagne}},
  \bibinfo{author}{\bibfnamefont{Y.}~\bibnamefont{Tarakanov}},
  \bibinfo{author}{\bibfnamefont{J.}~\bibnamefont{Kinaret}},
  \bibinfo{author}{\bibfnamefont{D.}~\bibnamefont{Garcia-Sanchez}},
  \bibnamefont{and} \bibinfo{author}{\bibfnamefont{A.}~\bibnamefont{Bachtold}},
  \bibinfo{journal}{Science} \textbf{\bibinfo{volume}{325}},
  \bibinfo{pages}{1107} (\bibinfo{year}{2009}).

\bibitem[{\citenamefont{Poot and van~der Zant}(2012)}]{Poot-review}
\bibinfo{author}{\bibfnamefont{M.}~\bibnamefont{Poot}} \bibnamefont{and}
  \bibinfo{author}{\bibfnamefont{H.~S.~J.} \bibnamefont{van~der Zant}},
  \bibinfo{journal}{Phys. Rep.} \textbf{\bibinfo{volume}{511}},
  \bibinfo{pages}{273} (\bibinfo{year}{2012}).

\bibitem[{\citenamefont{Bulaev et~al.}(2008)\citenamefont{Bulaev, Trauzettel,
  and Loss}}]{Bulaev}
\bibinfo{author}{\bibfnamefont{D.~V.} \bibnamefont{Bulaev}},
  \bibinfo{author}{\bibfnamefont{B.}~\bibnamefont{Trauzettel}},
  \bibnamefont{and} \bibinfo{author}{\bibfnamefont{D.}~\bibnamefont{Loss}},
  \bibinfo{journal}{Phys. Rev. B} \textbf{\bibinfo{volume}{77}},
  \bibinfo{pages}{235301} (\bibinfo{year}{2008}).

\bibitem[{\citenamefont{Churchill et~al.}(2009)\citenamefont{Churchill,
  Kuemmeth, Harlow, Bestwick, Rashba, Flensberg, Stwertka, Taychatanapat,
  Watson, and Marcus}}]{ChurchillPRL}
\bibinfo{author}{\bibfnamefont{H.~O.~H.} \bibnamefont{Churchill}},
  \bibinfo{author}{\bibfnamefont{F.}~\bibnamefont{Kuemmeth}},
  \bibinfo{author}{\bibfnamefont{J.}~\bibnamefont{Harlow}},
  \bibinfo{author}{\bibfnamefont{A.~J.} \bibnamefont{Bestwick}},
  \bibinfo{author}{\bibfnamefont{E.~I.} \bibnamefont{Rashba}},
  \bibinfo{author}{\bibfnamefont{K.}~\bibnamefont{Flensberg}},
  \bibinfo{author}{\bibfnamefont{C.~H.} \bibnamefont{Stwertka}},
  \bibinfo{author}{\bibfnamefont{T.}~\bibnamefont{Taychatanapat}},
  \bibinfo{author}{\bibfnamefont{S.~K.} \bibnamefont{Watson}},
  \bibnamefont{and} \bibinfo{author}{\bibfnamefont{C.~M.}
  \bibnamefont{Marcus}}, \bibinfo{journal}{Phys. Rev. Lett.}
  \textbf{\bibinfo{volume}{102}}, \bibinfo{pages}{166802}
  (\bibinfo{year}{2009}).

\bibitem[{\citenamefont{Flensberg and Marcus}(2010)}]{FlensbergMarcus}
\bibinfo{author}{\bibfnamefont{K.}~\bibnamefont{Flensberg}} \bibnamefont{and}
  \bibinfo{author}{\bibfnamefont{C.~M.} \bibnamefont{Marcus}},
  \bibinfo{journal}{Phys. Rev. B} \textbf{\bibinfo{volume}{81}},
  \bibinfo{pages}{195418} (\bibinfo{year}{2010}).

\bibitem[{\citenamefont{Ando}(2000)}]{Ando}
\bibinfo{author}{\bibfnamefont{T.}~\bibnamefont{Ando}}, \bibinfo{journal}{J.
  Phys. Soc. Jap.} \textbf{\bibinfo{volume}{69}}, \bibinfo{pages}{1757}
  (\bibinfo{year}{2000}).

\bibitem[{\citenamefont{Jeong and Lee}(2009)}]{Jeong}
\bibinfo{author}{\bibfnamefont{J.-S.} \bibnamefont{Jeong}} \bibnamefont{and}
  \bibinfo{author}{\bibfnamefont{H.-W.} \bibnamefont{Lee}},
  \bibinfo{journal}{Phys. Rev. B} \textbf{\bibinfo{volume}{80}},
  \bibinfo{pages}{075409} (\bibinfo{year}{2009}).

\bibitem[{\citenamefont{Izumida et~al.}(2009)\citenamefont{Izumida, Sato, and
  Saito}}]{Izumida}
\bibinfo{author}{\bibfnamefont{W.}~\bibnamefont{Izumida}},
  \bibinfo{author}{\bibfnamefont{K.}~\bibnamefont{Sato}}, \bibnamefont{and}
  \bibinfo{author}{\bibfnamefont{R.}~\bibnamefont{Saito}}, \bibinfo{journal}{J.
  Phys. Soc. Jap.} \textbf{\bibinfo{volume}{78}}, \bibinfo{pages}{074707}
  (\bibinfo{year}{2009}).

\bibitem[{\citenamefont{Rudner and Rashba}(2010)}]{RudnerRashba}
\bibinfo{author}{\bibfnamefont{M.~S.} \bibnamefont{Rudner}} \bibnamefont{and}
  \bibinfo{author}{\bibfnamefont{E.~I.} \bibnamefont{Rashba}},
  \bibinfo{journal}{Phys. Rev. B} \textbf{\bibinfo{volume}{81}},
  \bibinfo{pages}{125426} (\bibinfo{year}{2010}).

\bibitem[{\citenamefont{P\'{a}lyi et~al.}(2012)\citenamefont{P\'{a}lyi, Struck,
  Rudner, Flensberg, and Burkard}}]{Palyi-spinphonon}
\bibinfo{author}{\bibfnamefont{A.}~\bibnamefont{P\'{a}lyi}},
  \bibinfo{author}{\bibfnamefont{P.~R.} \bibnamefont{Struck}},
  \bibinfo{author}{\bibfnamefont{M.}~\bibnamefont{Rudner}},
  \bibinfo{author}{\bibfnamefont{K.}~\bibnamefont{Flensberg}},
  \bibnamefont{and} \bibinfo{author}{\bibfnamefont{G.}~\bibnamefont{Burkard}},
  \bibinfo{journal}{Phys. Rev. Lett.} \textbf{\bibinfo{volume}{108}},
  \bibinfo{pages}{206811} (\bibinfo{year}{2012}).

\bibitem[{\citenamefont{Lai et~al.}(2014)\citenamefont{Lai, Churchill, and
  Marcus}}]{Lai}
\bibinfo{author}{\bibfnamefont{R.~A.} \bibnamefont{Lai}},
  \bibinfo{author}{\bibfnamefont{H.~O.~H.} \bibnamefont{Churchill}},
  \bibnamefont{and} \bibinfo{author}{\bibfnamefont{C.~M.}
  \bibnamefont{Marcus}}, \bibinfo{journal}{Phys. Rev. B}
  \textbf{\bibinfo{volume}{89}}, \bibinfo{pages}{121303}
  (\bibinfo{year}{2014}).

\bibitem[{\citenamefont{Sz\'echenyi and P\'alyi}(2014)}]{Szechenyi-maximalrabi}
\bibinfo{author}{\bibfnamefont{G.}~\bibnamefont{Sz\'echenyi}} \bibnamefont{and}
  \bibinfo{author}{\bibfnamefont{A.}~\bibnamefont{P\'alyi}},
  \bibinfo{journal}{Phys. Rev. B} \textbf{\bibinfo{volume}{89}},
  \bibinfo{pages}{115409} (\bibinfo{year}{2014}).

\bibitem[{\citenamefont{Wang and Burkard}(2014)}]{Wang-control}
\bibinfo{author}{\bibfnamefont{H.}~\bibnamefont{Wang}} \bibnamefont{and}
  \bibinfo{author}{\bibfnamefont{G.}~\bibnamefont{Burkard}},
  \bibinfo{journal}{Phys. Rev. B} \textbf{\bibinfo{volume}{90}},
  \bibinfo{pages}{035415} (\bibinfo{year}{2014}).

\bibitem[{\citenamefont{Li et~al.}(2014)\citenamefont{Li, Benjamin, Briggs, and
  Laird}}]{Li-edsr}
\bibinfo{author}{\bibfnamefont{Y.}~\bibnamefont{Li}},
  \bibinfo{author}{\bibfnamefont{S.~C.} \bibnamefont{Benjamin}},
  \bibinfo{author}{\bibfnamefont{G.~A.~D.} \bibnamefont{Briggs}},
  \bibnamefont{and} \bibinfo{author}{\bibfnamefont{E.~A.} \bibnamefont{Laird}},
  \bibinfo{journal}{Phys. Rev. B} \textbf{\bibinfo{volume}{90}},
  \bibinfo{pages}{195440} (\bibinfo{year}{2014}).

\bibitem[{\citenamefont{Osika et~al.}(2014)\citenamefont{Osika,
  Mre\ifmmode~\acute{n}\else \'{n}\fi{}ca, and Szafran}}]{Osika-cnt}
\bibinfo{author}{\bibfnamefont{E.~N.} \bibnamefont{Osika}},
  \bibinfo{author}{\bibfnamefont{A.}~\bibnamefont{Mre\ifmmode~\acute{n}\else
  \'{n}\fi{}ca}}, \bibnamefont{and}
  \bibinfo{author}{\bibfnamefont{B.}~\bibnamefont{Szafran}},
  \bibinfo{journal}{Phys. Rev. B} \textbf{\bibinfo{volume}{90}},
  \bibinfo{pages}{125302} (\bibinfo{year}{2014}).

\bibitem[{\citenamefont{Ohm et~al.}(2012)\citenamefont{Ohm, Stampfer,
  Splettstoesser, and Wegewijs}}]{Ohm}
\bibinfo{author}{\bibfnamefont{C.}~\bibnamefont{Ohm}},
  \bibinfo{author}{\bibfnamefont{C.}~\bibnamefont{Stampfer}},
  \bibinfo{author}{\bibfnamefont{J.}~\bibnamefont{Splettstoesser}},
  \bibnamefont{and} \bibinfo{author}{\bibfnamefont{M.~R.}
  \bibnamefont{Wegewijs}}, \bibinfo{journal}{Appl. Phys. Lett.}
  \textbf{\bibinfo{volume}{100}}, \bibinfo{pages}{143103}
  (\bibinfo{year}{2012}).

\bibitem[{\citenamefont{Struck et~al.}(2014)\citenamefont{Struck, Wang, and
  Burkard}}]{Struck-readout}
\bibinfo{author}{\bibfnamefont{P.~R.} \bibnamefont{Struck}},
  \bibinfo{author}{\bibfnamefont{H.}~\bibnamefont{Wang}}, \bibnamefont{and}
  \bibinfo{author}{\bibfnamefont{G.}~\bibnamefont{Burkard}},
  \bibinfo{journal}{Phys. Rev. B} \textbf{\bibinfo{volume}{89}},
  \bibinfo{pages}{045404} (\bibinfo{year}{2014}).

\bibitem[{\citenamefont{Ono et~al.}(2002)\citenamefont{Ono, Austing, Tokura,
  and Tarucha}}]{Ono-spinblockade}
\bibinfo{author}{\bibfnamefont{K.}~\bibnamefont{Ono}},
  \bibinfo{author}{\bibfnamefont{D.~G.} \bibnamefont{Austing}},
  \bibinfo{author}{\bibfnamefont{Y.}~\bibnamefont{Tokura}}, \bibnamefont{and}
  \bibinfo{author}{\bibfnamefont{S.}~\bibnamefont{Tarucha}},
  \bibinfo{journal}{Science} \textbf{\bibinfo{volume}{297}},
  \bibinfo{pages}{1313} (\bibinfo{year}{2002}).

\bibitem[{\citenamefont{Jouravlev and Nazarov}(2006)}]{Jouravlev}
\bibinfo{author}{\bibfnamefont{O.~N.} \bibnamefont{Jouravlev}}
  \bibnamefont{and} \bibinfo{author}{\bibfnamefont{Y.~V.}
  \bibnamefont{Nazarov}}, \bibinfo{journal}{Phys. Rev. Lett.}
  \textbf{\bibinfo{volume}{96}}, \bibinfo{pages}{176804}
  (\bibinfo{year}{2006}).

\bibitem[{\citenamefont{Koppens et~al.}(2005)\citenamefont{Koppens, Folk,
  Elzerman, Hanson, van Beveren, Vink, Tranitz, Wegscheider, Kouwenhoven, and
  Vandersypen}}]{Koppens-spinblockade}
\bibinfo{author}{\bibfnamefont{F.~H.~L.} \bibnamefont{Koppens}},
  \bibinfo{author}{\bibfnamefont{J.~A.} \bibnamefont{Folk}},
  \bibinfo{author}{\bibfnamefont{J.~M.} \bibnamefont{Elzerman}},
  \bibinfo{author}{\bibfnamefont{R.}~\bibnamefont{Hanson}},
  \bibinfo{author}{\bibfnamefont{L.~H.~W.} \bibnamefont{van Beveren}},
  \bibinfo{author}{\bibfnamefont{T.}~\bibnamefont{Vink}},
  \bibinfo{author}{\bibfnamefont{H.~P.} \bibnamefont{Tranitz}},
  \bibinfo{author}{\bibfnamefont{W.}~\bibnamefont{Wegscheider}},
  \bibinfo{author}{\bibfnamefont{L.~P.} \bibnamefont{Kouwenhoven}},
  \bibnamefont{and} \bibinfo{author}{\bibfnamefont{L.~M.~K.}
  \bibnamefont{Vandersypen}}, \bibinfo{journal}{Science}
  \textbf{\bibinfo{volume}{309}}, \bibinfo{pages}{1346} (\bibinfo{year}{2005}).

\bibitem[{\citenamefont{Buitelaar et~al.}(2008)\citenamefont{Buitelaar,
  Fransson, Cantone, Smith, Anderson, Jones, Ardavan, Khlobystov, Watt,
  Porfyrakis et~al.}}]{Buitelaar}
\bibinfo{author}{\bibfnamefont{M.}~\bibnamefont{Buitelaar}},
  \bibinfo{author}{\bibfnamefont{J.}~\bibnamefont{Fransson}},
  \bibinfo{author}{\bibfnamefont{A.}~\bibnamefont{Cantone}},
  \bibinfo{author}{\bibfnamefont{C.}~\bibnamefont{Smith}},
  \bibinfo{author}{\bibfnamefont{D.}~\bibnamefont{Anderson}},
  \bibinfo{author}{\bibfnamefont{G.}~\bibnamefont{Jones}},
  \bibinfo{author}{\bibfnamefont{A.}~\bibnamefont{Ardavan}},
  \bibinfo{author}{\bibfnamefont{A.}~\bibnamefont{Khlobystov}},
  \bibinfo{author}{\bibfnamefont{A.}~\bibnamefont{Watt}},
  \bibinfo{author}{\bibfnamefont{K.}~\bibnamefont{Porfyrakis}},
  \bibnamefont{et~al.}, \bibinfo{journal}{Phys. Rev. B}
  \textbf{\bibinfo{volume}{77}}, \bibinfo{pages}{245439}
  (\bibinfo{year}{2008}).

\bibitem[{\citenamefont{Fransson and Rasander}(2006)}]{Fransson}
\bibinfo{author}{\bibfnamefont{J.}~\bibnamefont{Fransson}} \bibnamefont{and}
  \bibinfo{author}{\bibfnamefont{M.}~\bibnamefont{Rasander}},
  \bibinfo{journal}{Phys. Rev. B} \textbf{\bibinfo{volume}{73}},
  \bibinfo{pages}{205333} (\bibinfo{year}{2006}).

\bibitem[{\citenamefont{Chorley et~al.}(2011)\citenamefont{Chorley, Giavaras,
  Wabnig, Jones, Smith, Briggs, and Buitelaar}}]{Chorley}
\bibinfo{author}{\bibfnamefont{S.}~\bibnamefont{Chorley}},
  \bibinfo{author}{\bibfnamefont{G.}~\bibnamefont{Giavaras}},
  \bibinfo{author}{\bibfnamefont{J.}~\bibnamefont{Wabnig}},
  \bibinfo{author}{\bibfnamefont{G.}~\bibnamefont{Jones}},
  \bibinfo{author}{\bibfnamefont{C.}~\bibnamefont{Smith}},
  \bibinfo{author}{\bibfnamefont{G.}~\bibnamefont{Briggs}}, \bibnamefont{and}
  \bibinfo{author}{\bibfnamefont{M.}~\bibnamefont{Buitelaar}},
  \bibinfo{journal}{Phys. Rev. Lett.} \textbf{\bibinfo{volume}{106}},
  \bibinfo{pages}{206801} (\bibinfo{year}{2011}).

\bibitem[{\citenamefont{Petta et~al.}(2005)\citenamefont{Petta, Johnson,
  Taylor, Laird, Yacoby, Lukin, Marcus, Hanson, and Gossard}}]{Petta}
\bibinfo{author}{\bibfnamefont{J.~R.} \bibnamefont{Petta}},
  \bibinfo{author}{\bibfnamefont{A.~C.} \bibnamefont{Johnson}},
  \bibinfo{author}{\bibfnamefont{J.~M.} \bibnamefont{Taylor}},
  \bibinfo{author}{\bibfnamefont{E.~A.} \bibnamefont{Laird}},
  \bibinfo{author}{\bibfnamefont{A.}~\bibnamefont{Yacoby}},
  \bibinfo{author}{\bibfnamefont{M.~D.} \bibnamefont{Lukin}},
  \bibinfo{author}{\bibfnamefont{C.~M.} \bibnamefont{Marcus}},
  \bibinfo{author}{\bibfnamefont{M.~P.} \bibnamefont{Hanson}},
  \bibnamefont{and} \bibinfo{author}{\bibfnamefont{A.~C.}
  \bibnamefont{Gossard}}, \bibinfo{journal}{Science}
  \textbf{\bibinfo{volume}{309}}, \bibinfo{pages}{2180} (\bibinfo{year}{2005}).

\bibitem[{\citenamefont{Koppens et~al.}(2006)\citenamefont{Koppens, Buizert,
  Tielrooij, Vink, Nowack, Meunier, Kouwenhoven, and
  Vandersypen}}]{Koppens-esr}
\bibinfo{author}{\bibfnamefont{F.~H.~L.} \bibnamefont{Koppens}},
  \bibinfo{author}{\bibfnamefont{C.}~\bibnamefont{Buizert}},
  \bibinfo{author}{\bibfnamefont{K.~J.} \bibnamefont{Tielrooij}},
  \bibinfo{author}{\bibfnamefont{I.~T.} \bibnamefont{Vink}},
  \bibinfo{author}{\bibfnamefont{K.~C.} \bibnamefont{Nowack}},
  \bibinfo{author}{\bibfnamefont{T.}~\bibnamefont{Meunier}},
  \bibinfo{author}{\bibfnamefont{L.~P.} \bibnamefont{Kouwenhoven}},
  \bibnamefont{and} \bibinfo{author}{\bibfnamefont{L.~M.~K.}
  \bibnamefont{Vandersypen}}, \bibinfo{journal}{Nature}
  \textbf{\bibinfo{volume}{442}}, \bibinfo{pages}{766} (\bibinfo{year}{2006}).

\bibitem[{\citenamefont{Hanson et~al.}(2007)\citenamefont{Hanson, Kouwenhoven,
  Petta, Tarucha, and Vandersypen}}]{Hanson-rmp}
\bibinfo{author}{\bibfnamefont{R.}~\bibnamefont{Hanson}},
  \bibinfo{author}{\bibfnamefont{L.~P.} \bibnamefont{Kouwenhoven}},
  \bibinfo{author}{\bibfnamefont{J.~R.} \bibnamefont{Petta}},
  \bibinfo{author}{\bibfnamefont{S.}~\bibnamefont{Tarucha}}, \bibnamefont{and}
  \bibinfo{author}{\bibfnamefont{L.~M.~K.} \bibnamefont{Vandersypen}},
  \bibinfo{journal}{Rev. Mod. Phys.} \textbf{\bibinfo{volume}{79}},
  \bibinfo{pages}{1217} (\bibinfo{year}{2007}).

\bibitem[{\citenamefont{P\'{a}lyi and Burkard}(2009)}]{Palyi-hyperfine}
\bibinfo{author}{\bibfnamefont{A.}~\bibnamefont{P\'{a}lyi}} \bibnamefont{and}
  \bibinfo{author}{\bibfnamefont{G.}~\bibnamefont{Burkard}},
  \bibinfo{journal}{Phys. Rev. B} \textbf{\bibinfo{volume}{80}},
  \bibinfo{pages}{201404} (\bibinfo{year}{2009}).

\bibitem[{\citenamefont{P\'{a}lyi and Burkard}(2010)}]{Palyi-cnt-spinblockade}
\bibinfo{author}{\bibfnamefont{A.}~\bibnamefont{P\'{a}lyi}} \bibnamefont{and}
  \bibinfo{author}{\bibfnamefont{G.}~\bibnamefont{Burkard}},
  \bibinfo{journal}{Phys. Rev. B} \textbf{\bibinfo{volume}{82}},
  \bibinfo{pages}{155424} (\bibinfo{year}{2010}).

\bibitem[{\citenamefont{Weiss et~al.}(2010)\citenamefont{Weiss, Rashba,
  Kuemmeth, Churchill, and Flensberg}}]{Weiss}
\bibinfo{author}{\bibfnamefont{S.}~\bibnamefont{Weiss}},
  \bibinfo{author}{\bibfnamefont{E.}~\bibnamefont{Rashba}},
  \bibinfo{author}{\bibfnamefont{F.}~\bibnamefont{Kuemmeth}},
  \bibinfo{author}{\bibfnamefont{H.~O.~H.} \bibnamefont{Churchill}},
  \bibnamefont{and}
  \bibinfo{author}{\bibfnamefont{K.}~\bibnamefont{Flensberg}},
  \bibinfo{journal}{Phys. Rev. B} \textbf{\bibinfo{volume}{82}},
  \bibinfo{pages}{165427} (\bibinfo{year}{2010}).

\bibitem[{\citenamefont{von Stecher et~al.}(2010)\citenamefont{von Stecher,
  Wunsch, Lukin, Demler, and Rey}}]{vonStecher}
\bibinfo{author}{\bibfnamefont{J.}~\bibnamefont{von Stecher}},
  \bibinfo{author}{\bibfnamefont{B.}~\bibnamefont{Wunsch}},
  \bibinfo{author}{\bibfnamefont{M.}~\bibnamefont{Lukin}},
  \bibinfo{author}{\bibfnamefont{E.}~\bibnamefont{Demler}}, \bibnamefont{and}
  \bibinfo{author}{\bibfnamefont{A.~M.} \bibnamefont{Rey}},
  \bibinfo{journal}{Phys. Rev. B} \textbf{\bibinfo{volume}{82}},
  \bibinfo{pages}{125437} (\bibinfo{year}{2010}).

\bibitem[{\citenamefont{Reynoso and Flensberg}(2011)}]{Reynoso1}
\bibinfo{author}{\bibfnamefont{A.~A.} \bibnamefont{Reynoso}} \bibnamefont{and}
  \bibinfo{author}{\bibfnamefont{K.}~\bibnamefont{Flensberg}},
  \bibinfo{journal}{Phys. Rev. B} \textbf{\bibinfo{volume}{84}},
  \bibinfo{pages}{205449} (\bibinfo{year}{2011}).

\bibitem[{\citenamefont{Reynoso and Flensberg}(2012)}]{Reynoso2}
\bibinfo{author}{\bibfnamefont{A.~A.} \bibnamefont{Reynoso}} \bibnamefont{and}
  \bibinfo{author}{\bibfnamefont{K.}~\bibnamefont{Flensberg}},
  \bibinfo{journal}{Phys. Rev. B} \textbf{\bibinfo{volume}{85}},
  \bibinfo{pages}{195441} (\bibinfo{year}{2012}).

\bibitem[{\citenamefont{Kiss et~al.}(2011)\citenamefont{Kiss, P\'{a}lyi, Ihara,
  Wzietek, Alloul, Simon, Z\'{o}lyomi, Koltai, K\"{u}rti, D\'{o}ra
  et~al.}}]{Kiss}
\bibinfo{author}{\bibfnamefont{A.}~\bibnamefont{Kiss}},
  \bibinfo{author}{\bibfnamefont{A.}~\bibnamefont{P\'{a}lyi}},
  \bibinfo{author}{\bibfnamefont{Y.}~\bibnamefont{Ihara}},
  \bibinfo{author}{\bibfnamefont{P.}~\bibnamefont{Wzietek}},
  \bibinfo{author}{\bibfnamefont{H.}~\bibnamefont{Alloul}},
  \bibinfo{author}{\bibfnamefont{P.}~\bibnamefont{Simon}},
  \bibinfo{author}{\bibfnamefont{V.}~\bibnamefont{Z\'{o}lyomi}},
  \bibinfo{author}{\bibfnamefont{J.}~\bibnamefont{Koltai}},
  \bibinfo{author}{\bibfnamefont{J.}~\bibnamefont{K\"{u}rti}},
  \bibinfo{author}{\bibfnamefont{B.}~\bibnamefont{D\'{o}ra}},
  \bibnamefont{et~al.}, \bibinfo{journal}{Phys. Rev. Lett.}
  \textbf{\bibinfo{volume}{107}}, \bibinfo{pages}{187204}
  (\bibinfo{year}{2011}).

\bibitem[{\citenamefont{Sz\'echenyi and P\'alyi}(2013)}]{Szechenyihotspot}
\bibinfo{author}{\bibfnamefont{G.}~\bibnamefont{Sz\'echenyi}} \bibnamefont{and}
  \bibinfo{author}{\bibfnamefont{A.}~\bibnamefont{P\'alyi}},
  \bibinfo{journal}{Phys. Rev. B} \textbf{\bibinfo{volume}{88}},
  \bibinfo{pages}{235414} (\bibinfo{year}{2013}).

\bibitem[{\citenamefont{Danon et~al.}(2013)\citenamefont{Danon, Wang, and
  Manchon}}]{Danon-organic}
\bibinfo{author}{\bibfnamefont{J.}~\bibnamefont{Danon}},
  \bibinfo{author}{\bibfnamefont{X.}~\bibnamefont{Wang}}, \bibnamefont{and}
  \bibinfo{author}{\bibfnamefont{A.}~\bibnamefont{Manchon}},
  \bibinfo{journal}{Phys. Rev. Lett.} \textbf{\bibinfo{volume}{111}},
  \bibinfo{pages}{066802} (\bibinfo{year}{2013}).

\bibitem[{\citenamefont{Secchi and Rontani}(2013)}]{Secchi-intervalley}
\bibinfo{author}{\bibfnamefont{A.}~\bibnamefont{Secchi}} \bibnamefont{and}
  \bibinfo{author}{\bibfnamefont{M.}~\bibnamefont{Rontani}},
  \bibinfo{journal}{Phys. Rev. B} \textbf{\bibinfo{volume}{88}},
  \bibinfo{pages}{125403} (\bibinfo{year}{2013}).

\bibitem[{\citenamefont{Wunsch}(2009)}]{Wunsch}
\bibinfo{author}{\bibfnamefont{B.}~\bibnamefont{Wunsch}},
  \bibinfo{journal}{Phys. Rev. B} \textbf{\bibinfo{volume}{79}},
  \bibinfo{pages}{235408} (\bibinfo{year}{2009}).

\bibitem[{\citenamefont{Secchi and Rontani}(2009)}]{SecchiRontani}
\bibinfo{author}{\bibfnamefont{A.}~\bibnamefont{Secchi}} \bibnamefont{and}
  \bibinfo{author}{\bibfnamefont{M.}~\bibnamefont{Rontani}},
  \bibinfo{journal}{Phys. Rev. B} \textbf{\bibinfo{volume}{80}},
  \bibinfo{pages}{041404} (\bibinfo{year}{2009}).

\bibitem[{\citenamefont{Secchi and Rontani}(2010)}]{SecchiRontani2}
\bibinfo{author}{\bibfnamefont{A.}~\bibnamefont{Secchi}} \bibnamefont{and}
  \bibinfo{author}{\bibfnamefont{M.}~\bibnamefont{Rontani}},
  \bibinfo{journal}{Phys. Rev. B} \textbf{\bibinfo{volume}{82}},
  \bibinfo{pages}{035417} (\bibinfo{year}{2010}).

\bibitem[{\citenamefont{Ziani et~al.}(2013)\citenamefont{Ziani, Cavaliere, and
  Sassetti}}]{CavaliereWigner}
\bibinfo{author}{\bibfnamefont{N.~T.} \bibnamefont{Ziani}},
  \bibinfo{author}{\bibfnamefont{F.}~\bibnamefont{Cavaliere}},
  \bibnamefont{and} \bibinfo{author}{\bibfnamefont{M.}~\bibnamefont{Sassetti}},
  \bibinfo{journal}{J. Phys.: Condens. Matter} \textbf{\bibinfo{volume}{25}},
  \bibinfo{pages}{342201} (\bibinfo{year}{2013}).

\end{thebibliography}

\end{document}